\documentstyle[aps,pre,aps,graphicx]{revtex}

\newcommand{\T}{{\mathcal{T}}}
\newcommand{\sym}[1]{{\sf #1}} 

\renewcommand{\vec}[1]{{\bf #1}}
\title{Classical basis for quantum spectral fluctuations in hyperbolic systems}
\author{Sebastian M\"uller\\
{\it Fachbereich Physik, Universit\"at Duisburg-Essen, 45117 Essen, Germany}\\
E-mail: sebastian.mueller@uni-essen.de}
\begin{document}
\maketitle
\begin{abstract}
We reason in support of the universality of quantum spectral fluctuations in chaotic systems, starting from the pioneering work of Sieber and Richter who expressed the spectral form factor in terms of pairs of periodic orbits with self-crossings and avoided crossings. Dropping the restriction to uniformly hyperbolic dynamics, we show that for general hyperbolic two-freedom systems with time-reversal invariance the spectral form factor is faithful to random-matrix theory, up to quadratic order in time. We relate the action difference within the contributing pairs of orbits to properties of stable and unstable manifolds. In studying the effects of conjugate points, we show that almost self-retracing orbit loops do not contribute to the form factor. Our findings are substantiated by numerical evidence for the concrete example of two billiard systems.

\end{abstract}

\section{Introduction}
One of the fundamental questions of quantum chaos is why, in the semiclassical limit, almost all classically hyperbolic systems display universal spectral fluctuations, only depending on their symmetries. This universality was initially conjectured by Bohigas, Giannoni and Schmit \cite{bgs} and is by now supported by overwhelming experimental and numerical evidence \cite{haake,stoeckmann}.
Examples for experimental tests range from nuclear physics and atomic and molecular spectroscopy to classical microwave billiards studied in the limit of large wavenumbers.
Even though random-matrix theory provides a phenomenological description of these universal features, a derivation from first principles is still lacking. 
Also in other areas of research, such as mesoscopic quantum transport, the reasons for the amazing success of random-matrix theory are only beginning to emerge \cite{richtersieber}.
To tackle this long-lasting challenge, several approaches have been suggested such as parametric level dynamics \cite{haake} or an extension of field theoretical methods used in the theory of disordered systems \cite{disorder,wls}.

In the present paper, following an ansatz pioneered in \cite{berry,smilansky,siebrich}, we relate quantum spectral statistics to the properties of classical periodic orbits. We treat the spectral form factor ({\it i.e.} the Fourier transform of the 
spectral two-point correlation function). According to random-matrix theory, it has the following form for systems belonging to the orthogonal universality class ({\it i.e.} hyperbolic systems with no symmetries except a time-reversal symmetry whose time-reversal operator squares to unity)
\begin{eqnarray}
K(\tau)=2\tau-\tau\ln(1+2\tau)=2\tau-2\tau^2+2\tau^3\ldots.
\end{eqnarray}
Here, $\tau=\frac{T}{T_H}$ is the time measured in units of the Heisenberg time $T_H(E)=2\pi\hbar\bar{\rho}(E)$ (where $\bar{\rho}(E)$ is the average level density), and we are only considering the range $0<\tau<\frac{1}{2}$. 
Using Gutzwiller's trace formula for the level density \cite{gutzwiller}, the form factor can be expressed as a double sum over periodic orbits $\gamma$, $\gamma'$
\begin{eqnarray}
K(\tau)=\frac{1}{T_H}\left\langle\sum_{\gamma,\gamma'} A_\gamma A_{\gamma'}\exp\left(\frac{\rm i}{\hbar}(S_\gamma-S_{\gamma'})-{\rm i}(\mu_\gamma-\mu_{\gamma'})\frac{\pi}{2}\right)\delta\left(T-\frac{T_\gamma+T_{\gamma'}}{2}\right)\right\rangle_{E,T},\label{pairs}
\end{eqnarray}
where $A_\gamma$ is the classical stability amplitude, $T_\gamma$ the period and $\mu_\gamma$ the Maslov index of the periodic orbit $\gamma$; the brackets $\langle\ldots\rangle_{E,T}$ denote averaging over the energy and over a small time window. 
Note that as the semiclassical limit is taken for fixed $\tau$, the period of the contributing orbits tends to infinity proportionally to $T_H$.
The crucial point is that the summand connected to each pair of orbits has a phase given by their action difference divided by $\hbar$. Thus in the semiclassical limit, the phases will be randomly distributed and most terms will interfere destructively. A contribution to the form factor can only arise from pairs of orbits whose action difference is of the order of Planck's constant.
Thus, there is a deep relation between  correlations among the actions of \emph{classical periodic orbits} and correlations in \emph{quantum spectra} \cite{smilansky}.

First success in this direction was reported by Berry \cite{berry}, who derived the leading term $2\tau$ in the series expansion of $K(\tau)$ from pairs of orbits which are either identical or related by time reversal (diagonal approximation).
Starting with results by Argaman {\it et al.} \cite{smilansky}, the search was on to identify further families of orbit pairs with similar action, expected to give rise to higher-order contributions to the form factor (see \cite{smilansky2} and references therein).
A breakthrough was recently achieved by Sieber and Richter, who proposed orbit pairs in which one of the orbits contains a self-crossing in configuration-space with a small angle $\epsilon$ \cite{siebrich}. Its partner narrowly avoids that crossing, approximately following one loop of the first orbit, and following the other loop in the time-reversed sense (cp. Fig. \ref{fig:pair}). The action difference between the two is quadratic in $\epsilon$ and thus can be arbitrarily small. Since it is required that the time reversal of a classical orbit loop is again a classical orbit loop, such pairs of orbits can only exist in time-reversal invariant systems.
Using these pairs, Sieber and Richter derived the leading off-diagonal contribution $-2\tau^2$ to the form factor for a uniformly hyperbolic billiard (a billiard where all orbits have the same Lyapunov exponent $\lambda$), the so-called Hadamard-Gutzwiller model, {\it i.e.} geodesic motion on a tesselated surface of constant negative curvature of genus 2. They summed over the contributions of all these pairs using two ingredients, the action difference between the two partners and the density of crossing angles. The $\tau^2$-term arises due to a correction to the latter of next-to-leading order in the orbit period. It stems from the fact that in the Hadamard-Gutzwiller model, a loop with a small crossing angle must have a minimal traversal time $t_{\rm min}(\epsilon)=-\frac{2}{\lambda}\log c\epsilon$. In their derivation these authors made use of several system-specific niceties of the Hadamard-Gutzwiller model, which the form factor, being universal, cannot depend on.
First steps towards an extension to other systems have been taken for quantum graphs in \cite{berkolaiko,berkolaiko2}, where both the $\tau^2$- and the $\tau^3$-contribution to the spectral form factor were shown to originate from similar orbit pairs.

In the present paper, extending results we first presented in \cite{diplom,regensburg}, we go beyond these idealized models and derive the leading off-diagonal contribution to the form factor for general two-dimensional hyperbolic systems with a Hamiltonian of the form $H(\vec{Q},\vec{P})=\frac{\vec{P}^2}{2 m}+V(\vec{Q})$. The main novel ideas needed to establish this universality are:\\
(i) The relation between the partner orbits can be formulated elegantly in terms of the invariant manifolds, which also determine the action difference within the orbit pair.\\
(ii) The Maslov indices of the partner orbits can be shown to coincide.\\
(iii) In general systems, a logarithmic correction to the angle distribution arises involving the Lyapunov exponent of the system (as defined below).\\
(iv) In systems with conjugate points, the one-to-one correspondence between crossings and orbit pairs is broken.
There are crossings related to almost self-retracing loops without an associated partner orbit, and ``braids'' of crossings with a common partner. We show how to overcome these problems and reveal the distribution of crossings for which a partner orbit does exist as universal even in the presence of conjugate points.\\
(v) The universal contribution to the form factor follows from a relation between the invariant manifolds and the Lyapunov exponent.\\
Our findings are substantiated by numerical results for two billiard systems.

Note that the use of crossings in configuration space constitutes no conceptual problem {\it e.g.} concerning canonical invariance, as what we are using is in fact the geometry of the invariant manifolds.
As pointed out in \cite{nonconventional}, it appears natural to work in configuration space since it is singled out by the conventional time-reversal operator. Interestingly, the results of \cite{nonconventional} imply that our considerations immediately carry over to systems with non-conventional time-reversal invariance, as long as they can be canonically transformed to a Hamiltonian of the above structure. An alternative approach avoiding a projection to configuration space will be presented in \cite{spehner,turek}.

This paper is organized as follows. We first determine the action difference between the two partner orbits (Sect. 2) and show that their Maslov indices coincide (Sect. 3). In Section 4 we investigate the statistics of crossings angles, phase-space locations and loop times in systems without conjugate points. In Section 5 we generalize our findings to systems with conjugate points and clarify the relation between crossings and orbit pairs in such systems. Finally, we will derive the $\tau^2$-contribution to the form factor in Section 6.

\section{Action difference}
A long periodic orbit typically has a huge number of self-crossings in configuration space. For a given small-angle self-crossing, we will show that a related orbit with similar action exists, provided the two orbit loops separated by the crossing are long\footnote{Certain subtle issues concerning the existence and uniqueness of that partner will be dealt with in Section 6.}. This partner orbit is obtained from the initial orbit by time reversal of one loop and a local deformation close to the crossing. Further away, the deviation between the initial orbit (respectively its time-reversed) and the partner decays exponentially. As proposed in \cite{nonconventional}, the relation between these two orbits can be expressed elegantly in terms of stable and unstable manifolds. The main advantage of this method is that the action difference within the orbit pair can be shown to depend only on the crossing angle and the local behavior of the invariant manifolds at the phase-space location of the crossing. The complicated dependence on stability matrices of the two loops derived in \cite{siebrich} is thus drastically simplified in the limit of long loops.

Recall that two phase-space points lie on each other's stable (unstable) manifold if trajectories starting from these points come infinitely close for $t\rightarrow\infty$ ($t\rightarrow-\infty$). For two-freedom systems, the local behavior of these invariant manifolds at a phase-space point $\vec{X}$ can be characterized by just one number, called ``direction'' or ``curvature'' of the given manifold \cite{gaspard,review}. This direction is defined as the ratio between the momentum and configuration-space components $p$ and $q$ of an  infinitesimal phase-space deflection along the stable (unstable) manifold
\begin{eqnarray}
B_{s,u}(\vec{X})=\left.\frac{\partial p}{\partial q}\right|_{s,u}(\vec{X});
\end{eqnarray}
here, the deflection is assumed orthogonal to the trajectory passing through $\vec{X}$.
Note that time reversal $\T$ exchanges stable and unstable manifolds and changes the sign of their directions such that $B_s(\vec{X})=-B_u(\T\vec{X})$.

\begin{figure}\begin{center}\includegraphics*[scale=0.4]{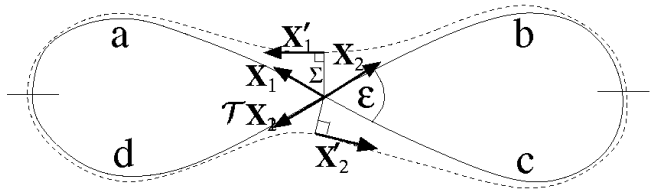}\end{center}\caption{Sketch of a Sieber-Richter pair in configuration space. 
Phase-space points are denoted by arrows starting at the corresponding configuration-space location, and momentum is indicated by the direction of the arrow. Depicted are
the phase-space points of the two traversals of the crossing and two phase-space points of the partner orbit as defined in the text. 
The Poincar\'e sections orthogonal to $\vec{X}_1'$ and $\vec{X}_2'$ and two points half-way through the loops divide the orbit into four parts $a$, $b$, $c$, and $d$.}\label{fig:pair}\end{figure}

Now, our condition of a local deformation can be formulated in terms of the invariant manifolds. 
Let $\vec{X}_1$ and $\vec{X}_2$ denote the phase-space points corresponding to the two traversals of the crossing. On the partner orbit with time-reversed right loop, we single out a phase-space point $\vec{X}_1'$, located where the perpendicular from the crossing hits that orbit (see Fig. \ref{fig:pair}). $\vec{X}_1'$ must approximately lie on the stable manifold of $\vec{X}_1$ (and vice versa), as trajectories starting there approach for a long time as $t\rightarrow\infty$.
Conversely, $\vec{X}_1'$ must lie on the unstable manifold of the time reversal of $\vec{X}_2$. To determine $\vec{X}_1'$, it is convenient to work in a Poincar\'e section $\Sigma$ defined by the above perpendicular. 
The coordinates of a phase-space point $\vec{X}=(\vec{Q},\vec{P})$ in that section will be denoted by $\vec{x}=(q,p)$.
Since $\vec{x}_1'-\vec{x}_1=(q_1'-q_1, p_1'-p_1)$ must be stable and $\vec{x}_1'-\T\vec{x}_2=(q_1'-q_2,p_1'+p_2)$ unstable, we obtain the following system of linear equations in $q_1'$, $p_1'$ involving the directions of the invariant manifolds
\begin{eqnarray}
p_1'-p_1&=&B_s(\vec{X}_1')(q_1'-q_1)\nonumber\\
p_1'+p_2&=&B_u(\vec{X}_1')(q_1'-q_2).\label{leq}
\end{eqnarray}
Furthermore, we know that $\vec{X}_1$ and $\vec{X}_2$ coincide in configuration space, {\it i.e.} also $q_1=q_2$, and that the difference in orthogonal momenta of $\vec{X}_1$ and $\T\vec{X}_2$ is related to the crossing angle by $p_1+p_2=-|\vec{P}_1|\epsilon$. Using this, the solution to (\ref{leq}) can be written as 
\begin{eqnarray}
q_1'-q_{1,2}=-\frac{|\vec{P}_1|\epsilon}{B_u(\vec{X}_1')-B_s(\vec{X}_1')},\label{ad1}
\end{eqnarray}
$p_1'$ follows trivially.

One can now expand the action of the orbit containing the crossing around that of its partner avoiding it. As the action of a periodic orbit is stationary, the first-order term vanishes. We restrict ourselves to the quadratic order in $q_1'-q_{1,2}$. 
We first consider the ``upper'' side of the crossing, {\it i.e.} the orbit parts $a$ and $b$ separated by $\Sigma$ and two points half-way through the two loops as shown in Figure \ref{fig:pair}.
The second derivatives of their action can be related to the stable and unstable directions. For example, $S_a(\vec{Q}_1',\vec{Q}_L')$ generates the motion from the configuration-space point $\vec{Q}_1'$ on $\Sigma$ to the point $\vec{Q}_L'$ half-way through the left loop, thus $\vec{P}_1'=-\frac{\partial S_a}{\partial\vec{Q}_1'}$ and therefore $\frac{\partial^2 S_a}{\partial q_1'^2}=-\frac{\partial p_1'}{\partial q_1'}$. These derivatives are taken for constant $\vec{Q}_L'$, and can in the limit of long loops be approximated by derivatives along the stable manifold. We thus have
\begin{eqnarray}
\frac{\partial^2S_a}{\partial {q_1'}^2}&=&-\left.\frac{\partial p_1'}{\partial q_1'}\right|_s(\vec{X}_1')=-B_s(\vec{X}_1'),
\end{eqnarray}
and analogously
\begin{eqnarray}
\frac{\partial^2S_b}{\partial {q_1'}^2}&=&\left.\frac{\partial p_1'}{\partial q_1'}\right|_u(\vec{X}_1')=B_u(\vec{X}_1').
\end{eqnarray}
The same reasoning can be repeated for the ``lower'' side of the crossing in a slightly different Poincar\'e section orthogonal to the ``lower'' part of the orbit in a phase-space point $\vec{X}_2'$ (see Fig. \ref{fig:pair}). 

A Taylor expansion now shows that avoiding a crossing
 with angle $\epsilon$ located at $\vec{X}$ reduces the action by
\begin{eqnarray}
\Delta S=\frac{\vec{P}_1^2\epsilon^2}{2(B_u(\vec{X}_1')-B_s(\vec{X}_1'))}+\frac{\vec{P}_1^2\epsilon^2}{2(B_u(\vec{X}_2')-B_s(\vec{X}_2'))}.\label{actiondiffsing}
\end{eqnarray}
For small angles, we can neglect the difference of the stable resp.
 unstable directions at $\vec{X}_1$, $\vec{X}_1'$, $\T\vec{X}_2$, and $\T\vec{X}_2'$ (which we will denote collectively by $\vec{X}$), as long as $B_s$ and $B_u$ are sufficiently smooth close to the crossing location. 
This condition is usually fulfilled unless {\it e.g.} immediately after the crossing, the different branches of
one loop enclose a singularity of the flow. 
Apart from these exceptional cases, (\ref{actiondiffsing}) simplifies to
\begin{eqnarray}
\Delta S(\vec{X},\epsilon)=\frac{\vec{P}^2\epsilon^2}{B_u(\vec{X})-B_s(\vec{X})}.\label{actiondiff}
\end{eqnarray}
This may also be negative, meaning that the action can also be increased by avoiding the crossing. Using that $B_s(\vec{X})=-B_u(\T\vec{X})$
one easily sees that this result is invariant under time reversal of $\vec{X}$, as it should be, since we must obtain the same action difference if we insert for $\vec{X}$ the other, almost time-reversed traversal of the crossing.
Assuringly, the action difference in the Hadamard-Gutzwiller model follows from (\ref{actiondiff}) as a special case. In that model we have $B_u(\vec{X})=-B_s(\vec{X})=m\lambda$ (where $m$ is mass) for all phase-space points $\vec{X}$, and thus $\Delta S=\frac{\vec{P}^2\epsilon^2}{2m\lambda}$ in accordance with \cite{siebrich}. 

\section{Maslov index}

The contribution of a pair of orbits to the form factor is determined not only by the difference of their actions, but also by the difference of their Maslov indices.
We want to show that for Sieber-Richter pairs of orbits, at least in the limit of small angles the latter vanishes. Note that this was trivial in the Hadamard-Gutzwiller model, where all Maslov indices are zero. 
We work in the framework of a beautiful geometric interpretation of the Maslov index of a periodic orbit due to Creagh {\it et al.} \cite{haake,littlejohn}, again in a (in the present context two-dimensional) Poincar\'e section orthogonal to the orbit. Here, the invariant manifolds locally have the form of lines through the origin which rotate around the origin as we move along the orbit. The Maslov index now equals the net number of clockwise half-rotations of the stable or, equivalently, the unstable manifold ({\it i.e.} the difference of the numbers of clockwise and counter-clockwise half-rotations).

For our argument, we also define the Maslov index of a finite non-periodic orbit such as an orbit loop in Sieber-Richter theory.
It is the sum of the net rotation angles of the stable and the unstable manifold around the origin, divided by $2\pi$. Even though not canonically invariant, non-integer, and depending on the units chosen in the Poincar\'e section, this definition is very useful because it makes the Maslov index of a loop invariant under time reversal. Obviously, time reversal leaves the absolute value of a rotation angle invariant. The same is true for the sense of rotation, since time reversal inverts the motion on the Poincar\'e section in direction (turning a clockwise rotation into a counter-clockwise one and vice versa), but also changes the sign of the momentum (turning the sense of rotation back to the original one)\footnote{Note that time reversal also inverts the directions of both coordinate axes in the Poincar\'e section, but this has no impact on the sense of rotation.}. An alternative analytic proof can be found in \cite{robbins}. In addition, time reversal exchanges the stable and unstable manifolds, which also cannot affect the \emph{sum} of their rotation angles. Reassuringly, this newly defined Maslov index coincides with the usual one in case the orbit happens to be periodic. Moreover, it is additive for subsequent orbit pieces and smooth under small deformations of the orbit as long as the invariant manifolds are smooth.

We can now show that the Maslov indices of the two partner orbits coincide. They can be expressed as sums over the Maslov indices of the two loops. Due to additivity and time-reversal invariance of the loop Maslov indices, formal time reversal of one loop leaves the Maslov index of the orbit invariant. In the limit of small crossing angles, any subsequent local deformation as described in Section 2 can at most lead to a small change of the Maslov indices of the loops. Since, however, the Maslov index of a periodic orbit is an integer quantity, the Maslov indices of the two partners have to coincide.

\section{Crossings in systems without conjugate points}

We have seen that in non-uniformly hyperbolic systems the action difference within a Sieber-Richter pair of orbits depends both on the angle and on the phase-space location of the crossing.
Anticipating that like in the Hadamard-Gutzwiller model, also the traversal times of the loops play a crucial role, we investigate the density $p(\vec{X},\epsilon,t|T)$ of loops with crossing angle $\epsilon$, time $t$, and initial phase-space point $\vec{X}$ (the latter being one of the two traversal points of the crossing) in a periodic orbit of period $T$.  
This density is normalized in such a way that integration over all possible values of $\vec{X}$, $\epsilon$, and $t$ yields the average number of loops, and thus twice the average number of crossings, in a periodic orbit of period close to $T$.
To make our exposition more clear, in the present Section we will still limit ourselves to systems without conjugate points; our findings will be generalized to systems with conjugate points in Section 5.

From the ergodicity of the flow we can, along the lines of \cite{siebrich}, deduce the following approximation
\begin{eqnarray}
p_{\rm erg}(\vec{X},\epsilon,t|T)=\frac{2\vec{P}^2}{m|\Omega|^2}T\sin\epsilon,
\end{eqnarray}
where $|\Omega|$ is the volume of the energy shell.
The proportionality to $\sin\epsilon$ reflects the fact that orthogonal parts of the orbit 
intersect more frequently than almost parallel or antiparallel ones.

However, like in the Hadamard-Gutzwiller model there is a correction to that ergodic prediction because loop times below a certain angle and location dependent minimum are impossible. We will first give a general argument why this minimal loop time is generic, and later substantiate our point by numerical evidence for billiards and discuss certain system-specific limitations. 
In systems without conjugate points, two trajectories starting from the same point in configuration space with a small opening angle cannot gather in another point as long as the phase-space separation between them can be approximated as a linear function of the initial separation. 
For an orbit loop to close it is thus necessary that, during half the loop time the separation between one of its two branches and the time-reversed of the other branch grows 
far enough to make that approximation invalid. Apart from exceptional cases to be discussed later, this typically means that their separation must reach a classical phase-space scale of the order of some finite fraction of the maximal phase-space separation.
If the branches enclose a small crossing angle $\epsilon$, this requires the loop to be long. Let $\|\delta\vec{X}\|$ denote the phase-space separation between one branch of the loop and the time-reversed of the other one in an arbitrary norm $\|.\|$ at the location of the crossing (thus $\|\delta\vec{X}\|\propto\epsilon$) and $\delta\vec{X}(\frac{t}{2})$ the separation after half the loop time. As long as the linearized approximation is applicable both are for large $t$ related by
\begin{eqnarray}
\frac{\|\delta\vec{X}(\frac{t}{2})\|}{\|\delta\vec{X}\|}\sim{\rm e}^{\frac{\lambda t}{2}},
\end{eqnarray}
where $\lambda$ is the Lyapunov exponent of the system. 
This asymptotic behavior is due to the fact that long loops ergodically explore the whole energy shell. The Lyapunov exponent governing the asymptotic fate of a small deviation $\delta\vec{X}$ (with non-vanishing unstable component) at a phase-space point $\vec{X}$ coincides for almost all $\vec{X}$, since it can be expressed as the average of the so-called local stretching rate over an infinite trajectory starting at $\vec{X}$ \cite{gaspard}. Thus, in case of an ergodic flow it coincides with a phase-space average almost everywhere (except {\it e.g.} on periodic orbits). Note that due to Pesin's theorem, $\lambda$ also coincides with the Kolmogorov-Sinai entropy \cite{gaspard,pesin}. In uniformly hyperbolic systems such averaging is trivial since the local stretching rate is constant and therefore even the Lyapunov exponents of periodic orbits coincide.

Demanding that the deviation $\|\delta\vec{X}(\frac{t}{2})\|$ reaches the limit for a breakdown of the linear approximation just after half the loop time, we obtain a minimal loop time of
\begin{eqnarray}
t_{\rm min}(\vec{X},\epsilon)=-\frac{2}{\lambda}\ln c(\vec{X})\epsilon\label{minloop}.
\end{eqnarray}
Here, $c(\vec{X})$ is constant with respect to the angle but may depend on the phase-space point $\vec{X}$ immediately preceding the loop.
However, the exact value of $c(\vec{X})$ will turn out to be irrelevant for the form factor. The second loop starts approximately at $\T\vec{X}$, thus its time must fulfill $T-t>t_{\rm min}(\T\vec{X},\epsilon)$. Incorporating these minimal loop times in a straight-forward way, we obtain the following density of loops
\begin{eqnarray}
p(\vec{X},\epsilon,t|T)=p_{\rm erg}(\vec{X},\epsilon,t|T)\Theta(t-t_{\rm min}(\vec{X},\epsilon))\Theta(T-t-t_{\rm min}(\T\vec{X},\epsilon))\label{anglelocal}\label{angledist}.
\end{eqnarray}
Integration over $0\leq t\leq T$ yields the density of crossing angles and phase-space locations\begin{eqnarray}
P(\vec{X},\epsilon|T)=
\frac{2\vec{P}^2}{m|\Omega|^2}T(T-t_{\rm min}(\vec{X},\epsilon)-t_{\rm min}(\T\vec{X},\epsilon))\sin\epsilon.\label{angle}
\end{eqnarray}

We want to clearly point out the scope of our approximation. First, the minimal loop time derived above gives the threshold after which loops have a \emph{chance} to close, but does not predict the exact behavior of the shortest loops. Thus in the immediate vicinity of the minimal loop time, system specific structures appear which are not described by (\ref{angledist}) and (\ref{angle}). Second, we will see that singularities can let the linear approximation for the separation between the branches of a loop break down even before a typical phase-space scale is reached (as also discussed in \cite{spehner}). Thus, singularities give rise to exceptional, system-specific crossings which may ignore the minimal loop time \cite{diplom}.

For our numerics, we are also interested in the statistics of self-crossings of non-periodic orbits. 
In full analogy to the above considerations for periodic orbits, we can show that the density of loops with crossing angle $\epsilon$, time $t$, and initial phase-space point $\vec{X}$ in non-periodic orbits with traversal time $T$ reads \begin{eqnarray}
p^{{\rm np}}(\vec{X},\epsilon,t|T)=
\frac{2\vec{P}^2}{m|\Omega|^2}(T-t)\Theta(t-t_{{\rm min}}(\vec{X},\epsilon))\sin\epsilon\label{np1}.
\end{eqnarray}
Integration yields the density of crossing angles and phase-space locations\footnote{Note that in contrast to periodic orbits, for each crossing there is only one loop, hence it is only included once in this distribution.}
\begin{eqnarray}
P^{{\rm np}}(\vec{X},\epsilon|T)=\frac{\vec{P}^2}{m|\Omega|^2}(T-t_{{\rm min}}(\vec{X},\epsilon))^2\sin\epsilon.\label{pnp}
\end{eqnarray}

\subsection{Example: The desymmetrized diamond billiard}

\begin{figure}\begin{center}\includegraphics*[scale=0.3]{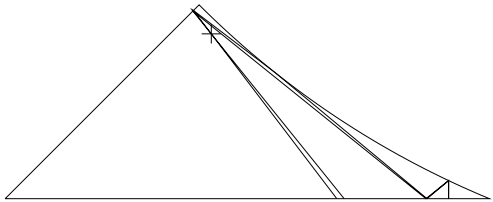}\end{center}\caption{The desymmetrized diamond billiard (with an example for a singularity-related crossing).}\label{fig:diamsing}\end{figure}

We will now present numerical evidence for this crossing distribution for a special billiard system, the desymmetrized diamond billiard (see Fig. \ref{fig:diamsing}).
It can be regarded as the empty space between four overlapping disks \cite{diamond} \`a la Sinai, cut into eight equal pieces. The distance between the disks is chosen as one, and we choose their radius $r=0.541$ so that 
its interior angles become $\frac{\pi}{2}$, $\frac{\pi}{4}$ and $\frac{\pi}{8}$. 
The desymmetrized diamond billiard is non-uniformly hyperbolic and belongs to the class of semi-dispersing billiards ({\it i.e.} it is surrounded by a boundary which consists of locally concave and straight segments). Therefore, it is free of conjugate points.
It has a circumference $C=0.671$ and an area $A=0.0157$. Santal\'o's formula \cite{mfp} gives its mean free path as $\bar{l}=\frac{\pi A}{C}=0.0735$. By averaging over the Lyapunov exponents of random non-periodic trajectories, we numerically obtain the Lyapunov exponent of the system as $\lambda = 4.31$.

A few words are in order about our numerical technique. The crossing statistics is determined by averaging over $2\times 10^7$ non-periodic trajectories of length $L=10$ ({\it i.e.} very long orbits compared to the typical length scales $\bar{l}$ and $\frac{1}{\lambda}$) with random initial conditions. For the delicate statistics of angles $<\frac{\pi}{10}$, even $10^{9}$ trajectories were considered. Note that for billiards, it is useful to work in dimensionless coordinates with mass and velocity equal to one; 
then the traversal time $T$ of each orbit equals $L$, and the same holds true for the the traversal times $t$ and lengths $l$ of the loops.

\begin{figure}\begin{center}\includegraphics[scale=0.34]{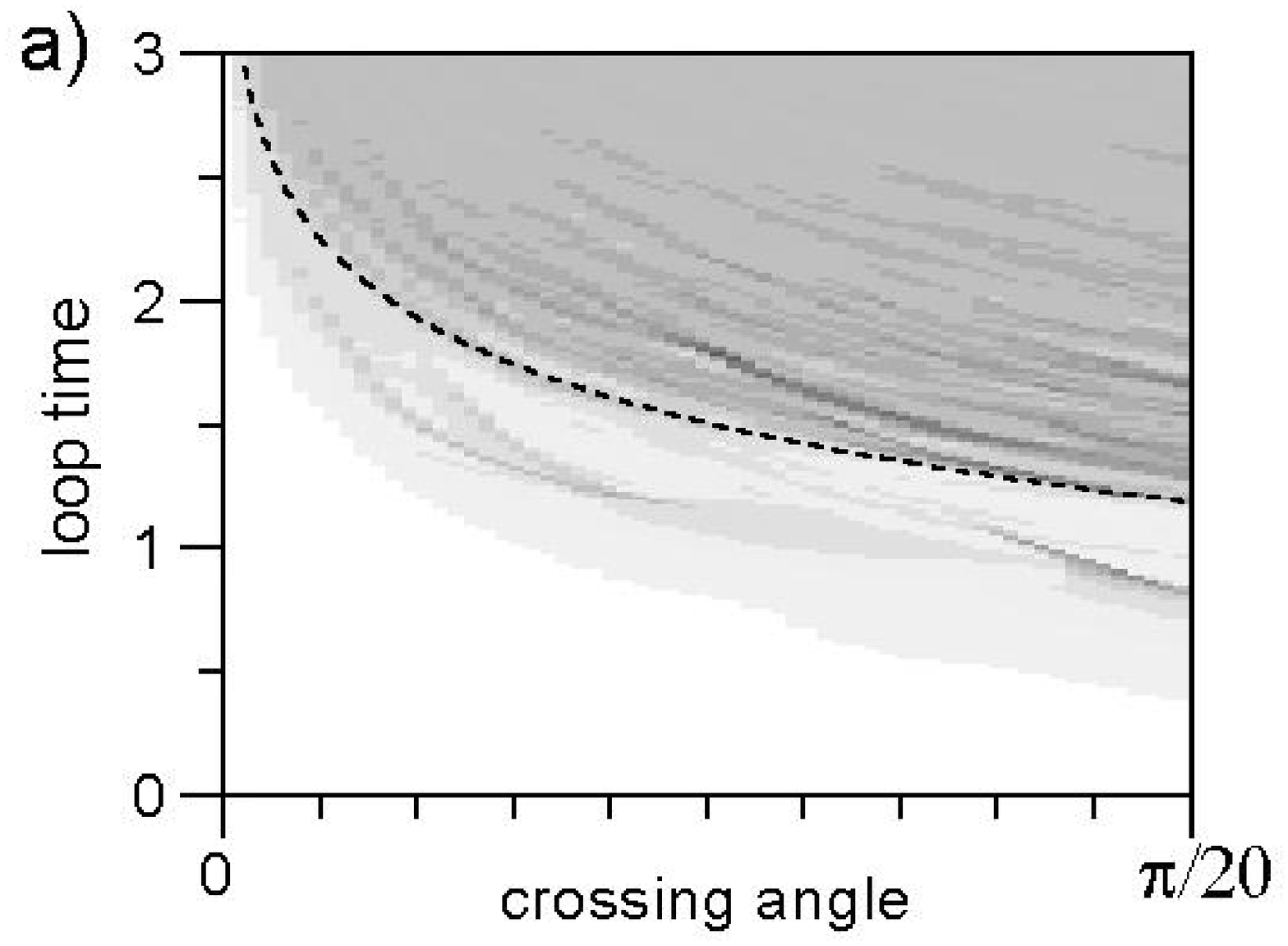}\hspace{1cm}\includegraphics[scale=0.34]{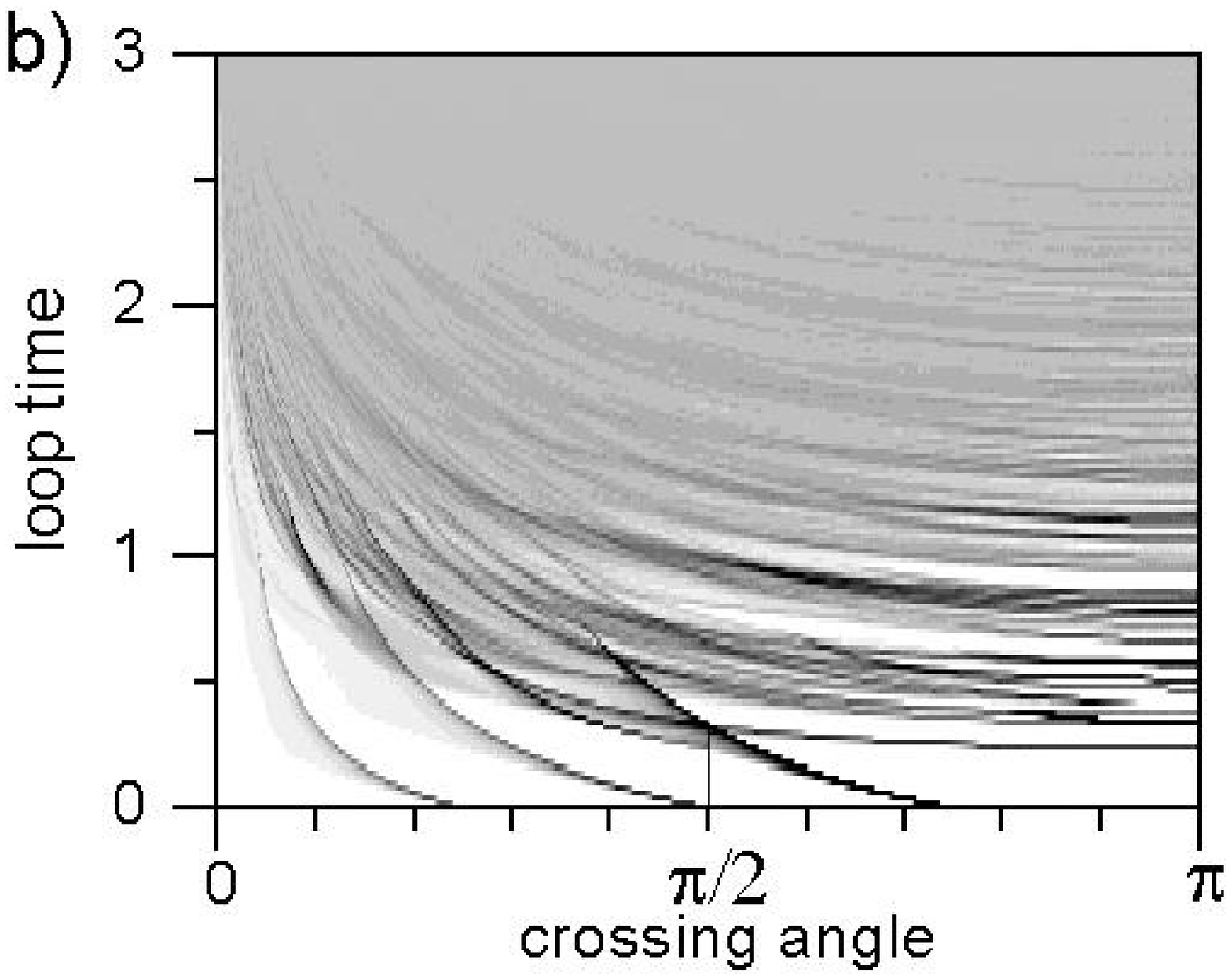}\hspace{1cm}\includegraphics[scale=0.17]{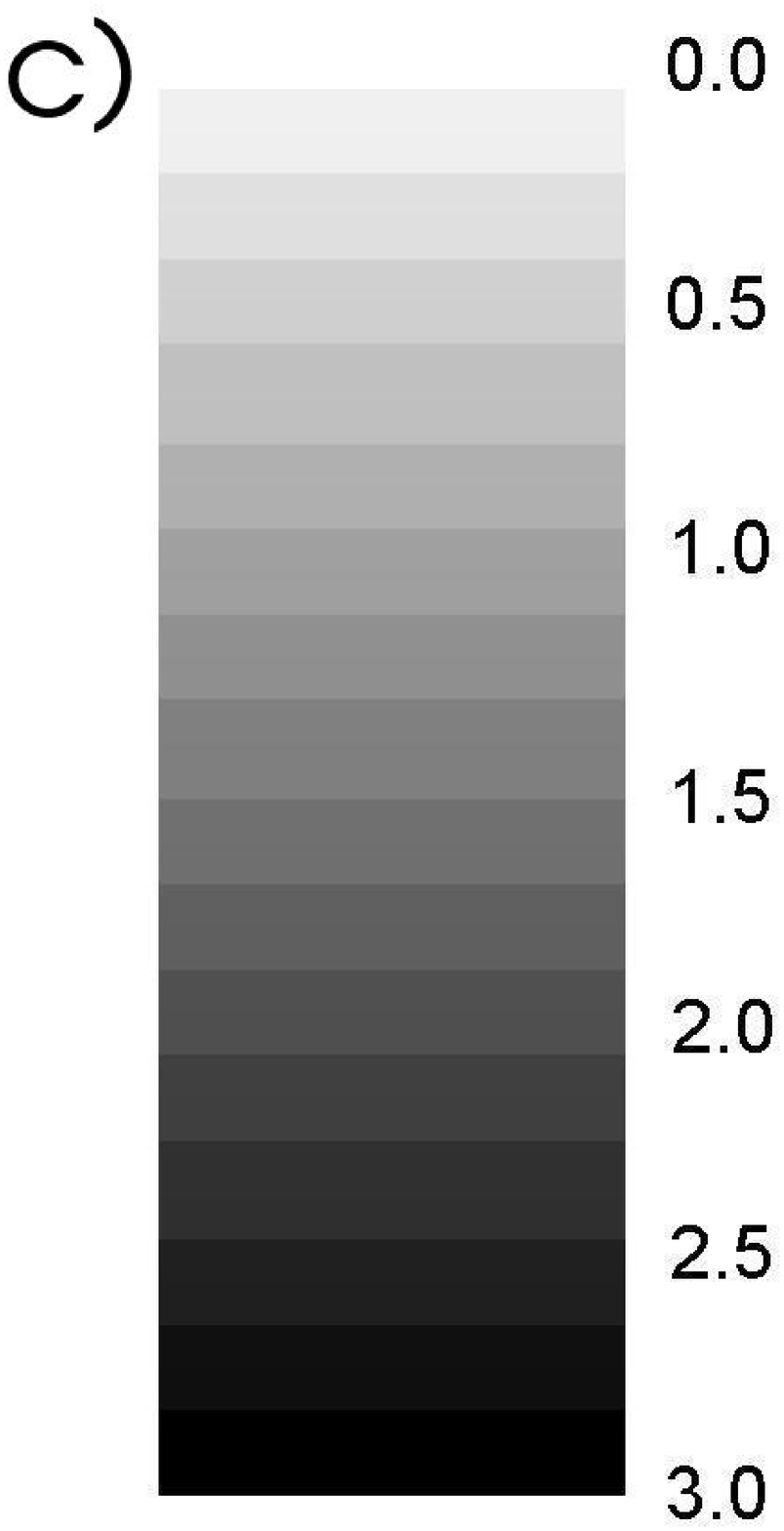}
\end{center}\caption{Density plot of the combined distribution of loop times $t$ and crossing angles $\epsilon$ in the desymmetrized diamond billiard: a) for $\epsilon<0.05\pi$, b) for all angles.
The density is normalized such that multiplied by $\sin\epsilon\hspace{0.05cm}d\epsilon\hspace{0.05cm}dt$, it gives the number of crossings in the respective intervals of the crossing angle $\epsilon$ and the loop time $t$  
inside one orbit of traversal time $T=10$; the resulting scale is shown in c).
For small angles, we observe a threshold logarithmic in $\epsilon$, as indicated by a dashed line.}\label{fig:diama}\end{figure}

We now turn to our results. For small angles, the basic idea of a minimal loop time depending logarithmitically on the crossing angle can be verified at a glance at Figure \ref{fig:diama}a. Here we have plotted  the density of angles and loop times (up to $t=3$) of crossings occurring anywhere on the energy shell\footnote{In contrast to the following results, for this density plot it was sufficient to only take into account $5\times10^7$ trajectories.}, and for convenience divided out the term $\sin\epsilon$ arising from the ergodic approximation. Sufficiently far above that logarithmic threshold, marked by a dashed line in Figure \ref{fig:diama}a, this density is almost uniform. In agreement with (\ref{np1}), it decays linearly towards larger loop time. 
Note that according to (\ref{minloop}), the minimal loop time weakly depends on the location on the energy shell due to the factor $c(\vec{X})$. Thus, the minimum in Figure \ref{fig:diama}a is slightly smeared out.
Below the dashed line, the density of crossings diminishes fast before vanishing completely inside the white region.

In addition to this expected behavior, a rich variety of system-specific structures are seen. 
As announced, the crossing density shows system-specific inhomogeneities in the immediate vicinity of the minimal loop time. In addition, below the minimal loop time, we observe exceptional crossings related to singularities of the flow, most importantly the tangential singularity. Namely, the linear approximation for the separation between the two branches of a loop is already violated if one branch reflects {\it e.g.} at the circular part of the boundary, and the other branch narrowly avoids the circle. In this case, the separation between the branches of the loop need not reach a typical phase space scale for the loop to close. An example for such exceptional loops ignoring the minimal loop time is shown in Figure \ref{fig:diamsing}. However, our numerical results indicate that their effect on the crossing distribution is minute.

At larger angles, 
which in the semiclassical limit give no contribution to the form factor, further structures are seen (cp. Fig. \ref{fig:diama}b), such as\\
(i) Discrete lines at $\epsilon=\pi$ corresponding to periodic orbits, which are nothing but loops with a ``crossing angle'' $\pi$. They are deformed and broadened when going to smaller angles. Thus, most loops close to the minimal time are obtained by deformation of the shortest periodic orbits.\\
(ii) Four families of loops which start at zero loop time, and as shown in \cite{diplom} are close to corners. 
One of these families of loops involves only reflections at the two straight-line segments of the boundary and has a crossing angle of $\frac{\pi}{2}$ for all its members.

Note that most of these structures were absent in the idealized example of the Hadamard-Gutzwiller model. In that model, the analog of Figure \ref{fig:diama} just consists of dispersionless logarithmic curves corresponding each to the family of loops obtained by deformation of one periodic orbit \cite{siebrich}.

\begin{figure}
\begin{center}
\begin{tabular}{lp{0.5cm}l}
\includegraphics*[scale=0.2]{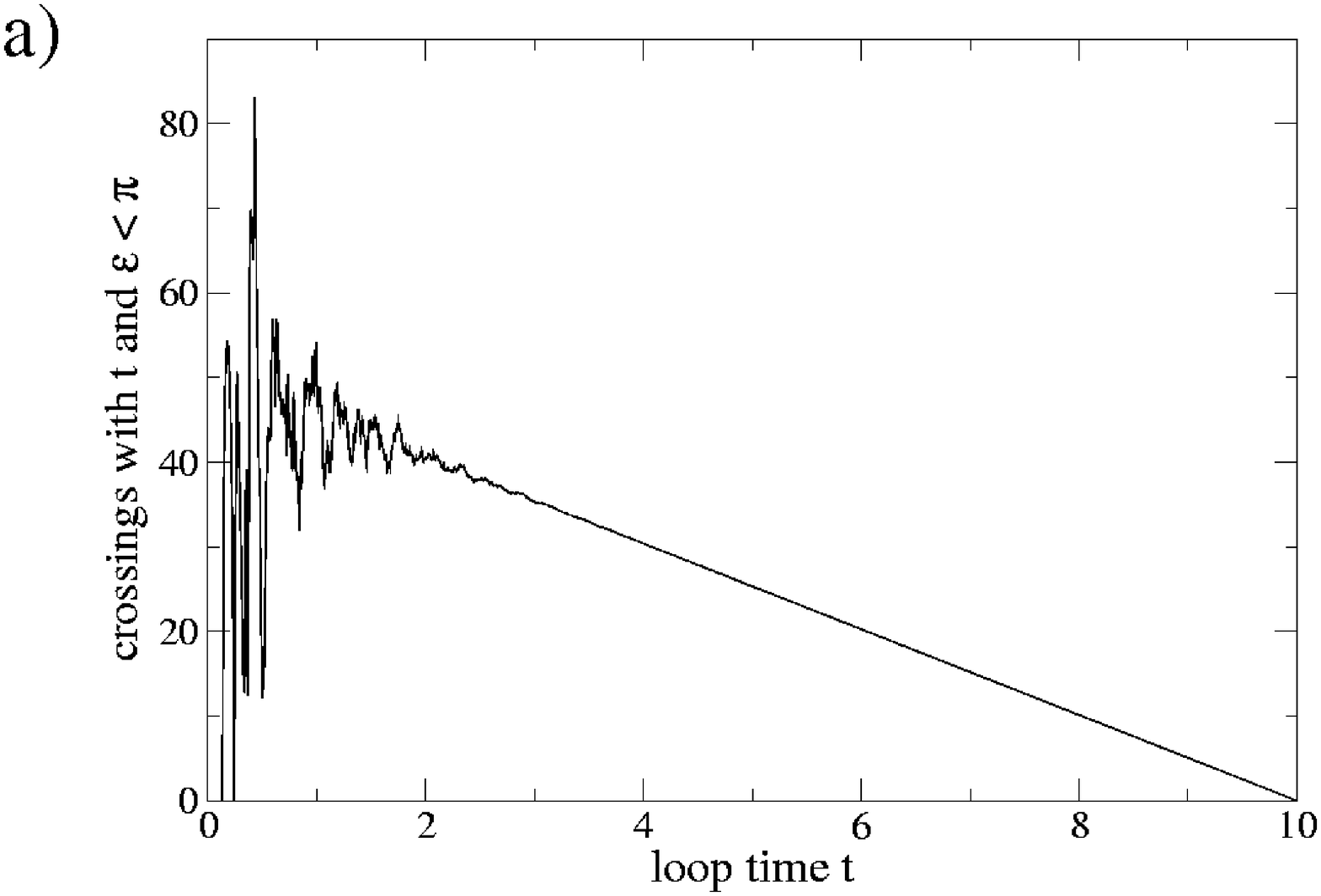}&&
\includegraphics*[scale=0.2]{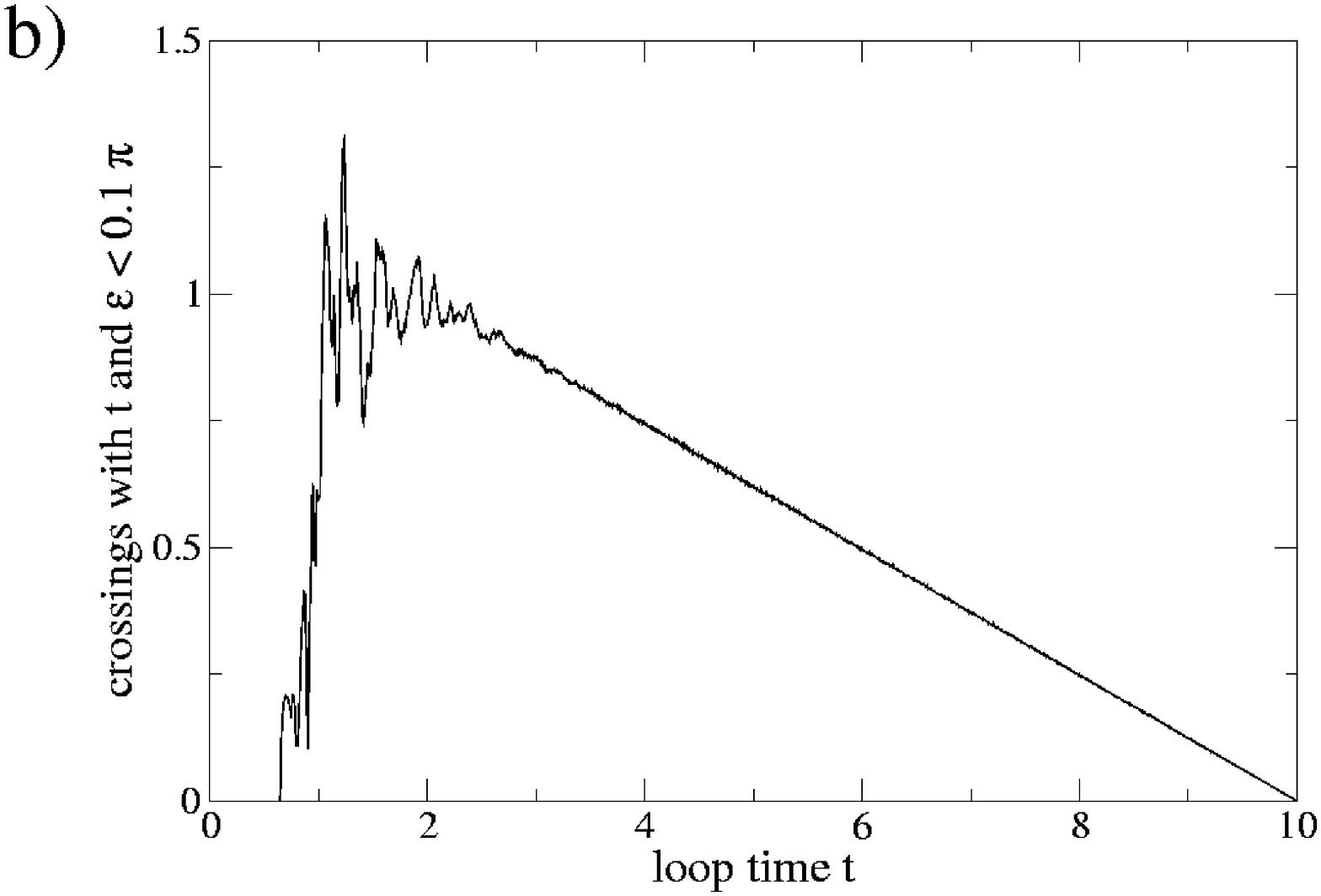}\\&&\\
\includegraphics*[scale=0.2]{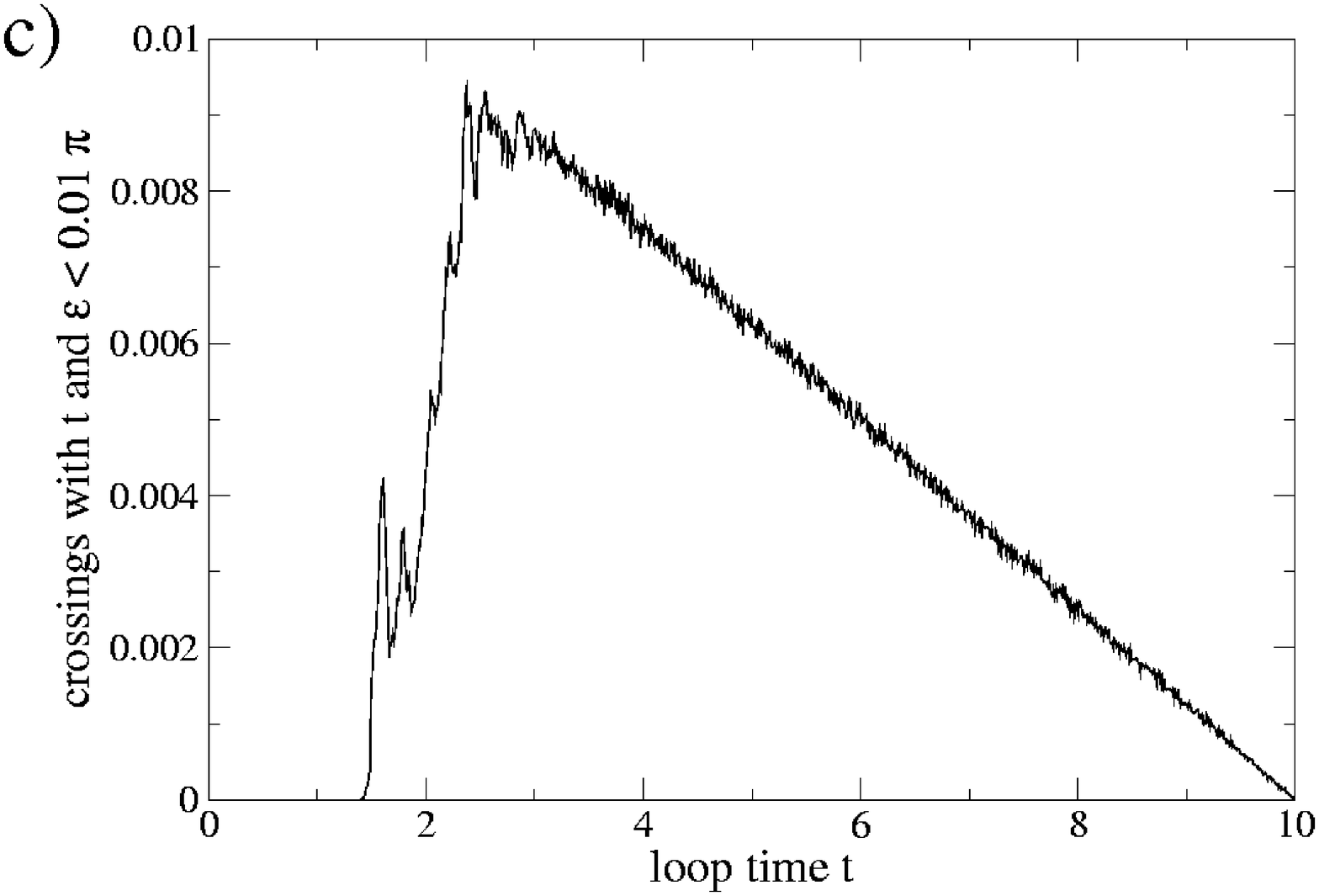}&&
\includegraphics*[scale=0.2]{l001.eps}
\end{tabular}
\end{center}
\caption{Statistics of loop times in non-periodic trajectories of traversal time $T=10$ in the desymmetrized diamond billiard.
Depicted is the
density of crossings with loop time $t$ and crossing angle $<\epsilon_{\rm max}$ in a fraction of $\frac{1}{8}$ of the energy shell for a) $\epsilon_{\rm max}=\pi$, b) $\epsilon_{\rm max}=0.1\pi$, c) $\epsilon_{\rm max}=0.01\pi$, d) $\epsilon_{\rm max}=0.001\pi$. 
The result is normalized such that multiplied by $dt$ it gives the number of such crossings in the loop time interval $(t,t+dt)$ inside one orbit of traversal time $T$; the graphs are based on averages over $2\times 10^7$ orbits for large angles and $10^9$ orbits for small angles, respectively.
Due to the minimal loop time, there is a gap for small $t$ which grows as $\epsilon_{\rm max}$ is reduced.}
\label{fig:l10}
\end{figure}

Further evidence for the minimal loop time is shown in Figure \ref{fig:l10}, this time also taking into account the phase-space dependence of the crossing statistics. We restrict ourselves to crossings which have angles smaller than a maximal angle $\epsilon_{\rm max}$ and take place inside a bin whose area is $\frac{1}{8}$ of the total energy shell, and consider the density of their loop times $t$. This density must have a gap for small $t$ whose width is given by the minimal loop time and thus increases logarithmitically when $\epsilon_{\rm max}$ is reduced. Incidentally, this density of loop times can be shown to decay linearly for large $t$ in the case of non-periodic orbits. Our numerical results confirm these predictions and again reveal system-specific structures close to the minimal loop time.

\begin{figure}
\begin{center}
\includegraphics*[scale=0.25]{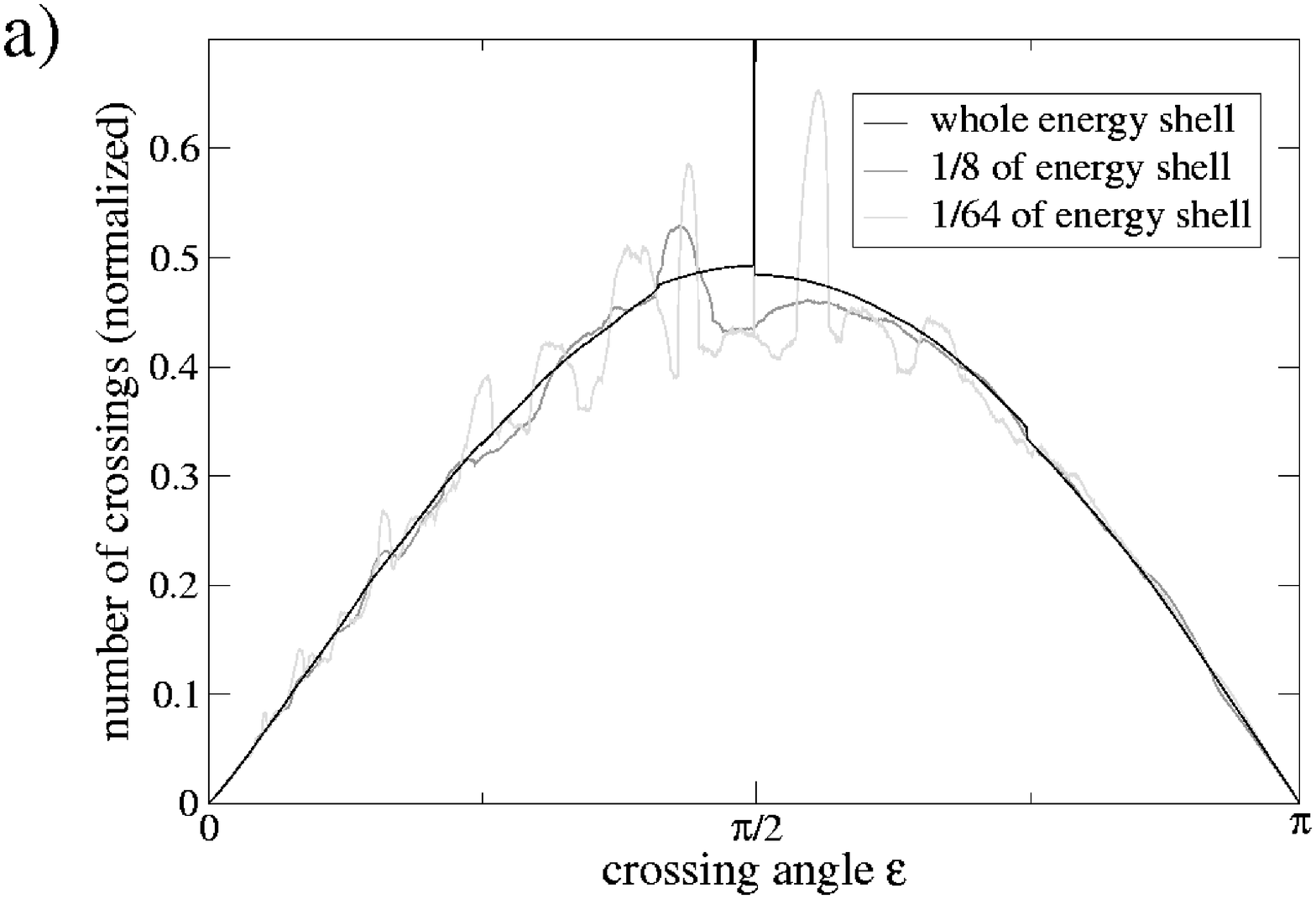}\hspace{1cm}
\includegraphics*[scale=0.25]{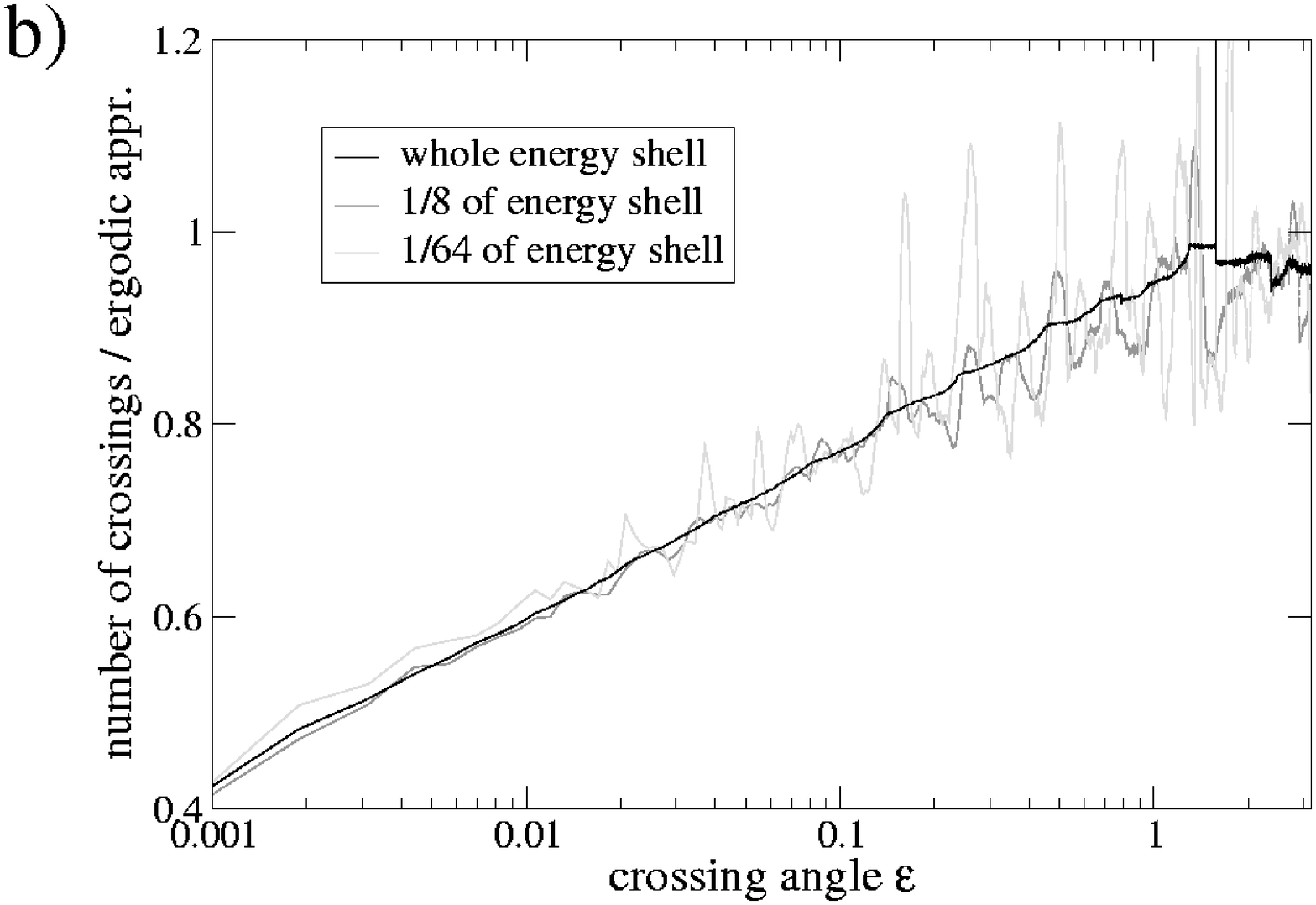}
\end{center}
\caption{a) Average of $P^{\rm np}(\vec{X},\epsilon|T)$ over the whole energy shell of the desymmetrized diamond billiard and fractions of $\frac{1}{8}$ and $\frac{1}{64}$ thereof (divided by $\frac{2\vec{P}^2T^2}{m|\Omega|^2}$ for normalization). b) The same quantities divided by $\frac{\sin\epsilon}{2}$ in a logarithmic plot. The logarithmic correction due to the minimal loop time becomes visible.}\label{fig:angle}
\end{figure}

The angle density $P^{\rm np}(\vec{X},\epsilon|T)$ (averaged over the whole energy shell and over bins with sizes of $\frac{1}{8}$ and $\frac{1}{64}$ of the energy shell) agrees with our predictions as well (cp. Fig. \ref{fig:angle}). For sufficiently small angles, the distribution is sinusoidal. Upon division by $\sin\epsilon$, in accordance with (\ref{pnp}) a logarithmic correction due to the minimal loop time becomes visible. The relative weight of that correction (seen as the slope in the logarithmic plot in Fig. \ref{fig:angle}b) coincides for all three samples. We thus see that indeed the crossing distribution in a billiard is homogeneous on the energy shell apart from system-specific oscillations for large $\epsilon$ and apart from the factor $c(\vec{X})$. 
Based on the relative weight of the logarithm, fitting yields a minimal loop time $t_{\rm min}(\vec{X},\epsilon)=-\frac{2}{\lambda_{\rm fit}}\ln c(\vec{X})\epsilon$ where $\lambda_{\rm fit}=4.31$ agrees perfectly with the Lyapunov exponent $\lambda$.  

\section{Crossings in systems with conjugate points}

\begin{figure}\begin{center}\includegraphics*[scale=0.27]{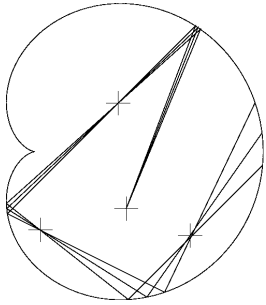}\end{center}\caption{Example for a family of mutually conjugate points in the cardioid billiard.}\label{fig:conjugate}\end{figure}

We now want to generalize our treatment to systems with conjugate points. In such systems, trajectories fanning out from the same point in configuration space with a small opening angle can focus again in a second point, then in a third, etc. (see Fig. \ref{fig:conjugate}). All these points are called mutually conjugate. We will see that conjugate points destroy the one-to-one relation between crossings and orbit pairs, because there are (i) crossings without a partner orbit avoiding the crossing and (ii) families (``braids'') of crossings with a common partner. 
Both effects have a direct analogy to quantum graphs, for which the leading off-diagonal contribution to the form factor was derived in \cite{berkolaiko} and even the third order of the expansion was obtained \cite{berkolaiko2}.
The general picture emerges that a periodic orbit has one partner for each two almost time-reversed orbit stretches dividing the orbit into sufficiently long loops.

Even though our findings are general, we find it instructive to also discuss their meaning in the special case of systems with symbolic dynamics. Here, each periodic orbit is unambiguously defined by a string of symbols. Each symbol in the alphabet denotes one partition of a certain Poincar\'e section, and the symbol sequence of a periodic orbit is composed of the symbols corresponding to the partitions it traverses.

In particular, we consider the example of the cardioid billiard, which belongs to the family of focusing billiards ({\it i.e.} billiards surrounded by a locally convex boundary). The cardioid has been intensively studied in the literature, see \cite{robnik,baeckerspec,bruuswhelan,baeckerdullin,baeckerchernov,baeckerdiss} and references therein. 
It is hyperbolic, and the fidelity of both symmetry-reduced spectra to random-matrix theory was demonstrated numerically in \cite{baeckerspec}.
It has symbolic dynamics with two symbols effectively denoting straight-line segments of the orbit \cite{bruuswhelan,baeckerdullin,baeckerchernov}. The initial points of each segment and the cusp divide the boundary into two parts.
A symbol $\sym{A}$ is assigned to the segment if, seen from the initial point, the final one lies in the part on the clockwise side, and a symbol $\sym{B}$ if it lies on the counter-clockwise side. We note that time reversal inverts the ordering of symbols and interchanges $\sym{A}$ and $\sym{B}$.
For a more detailed account of the following results, we refer the reader to \cite{diplom}.

\subsection{Crossings without partner}

In systems with conjugate points, loops below the minimal loop time derived above are possible, because the two branches of a loop can meet while the linear approximation for their separation is still applicable. However, we will see that 
in this case the orbit has no partner avoiding the crossing.

\begin{figure}
\begin{center}
\includegraphics*[scale=0.27]{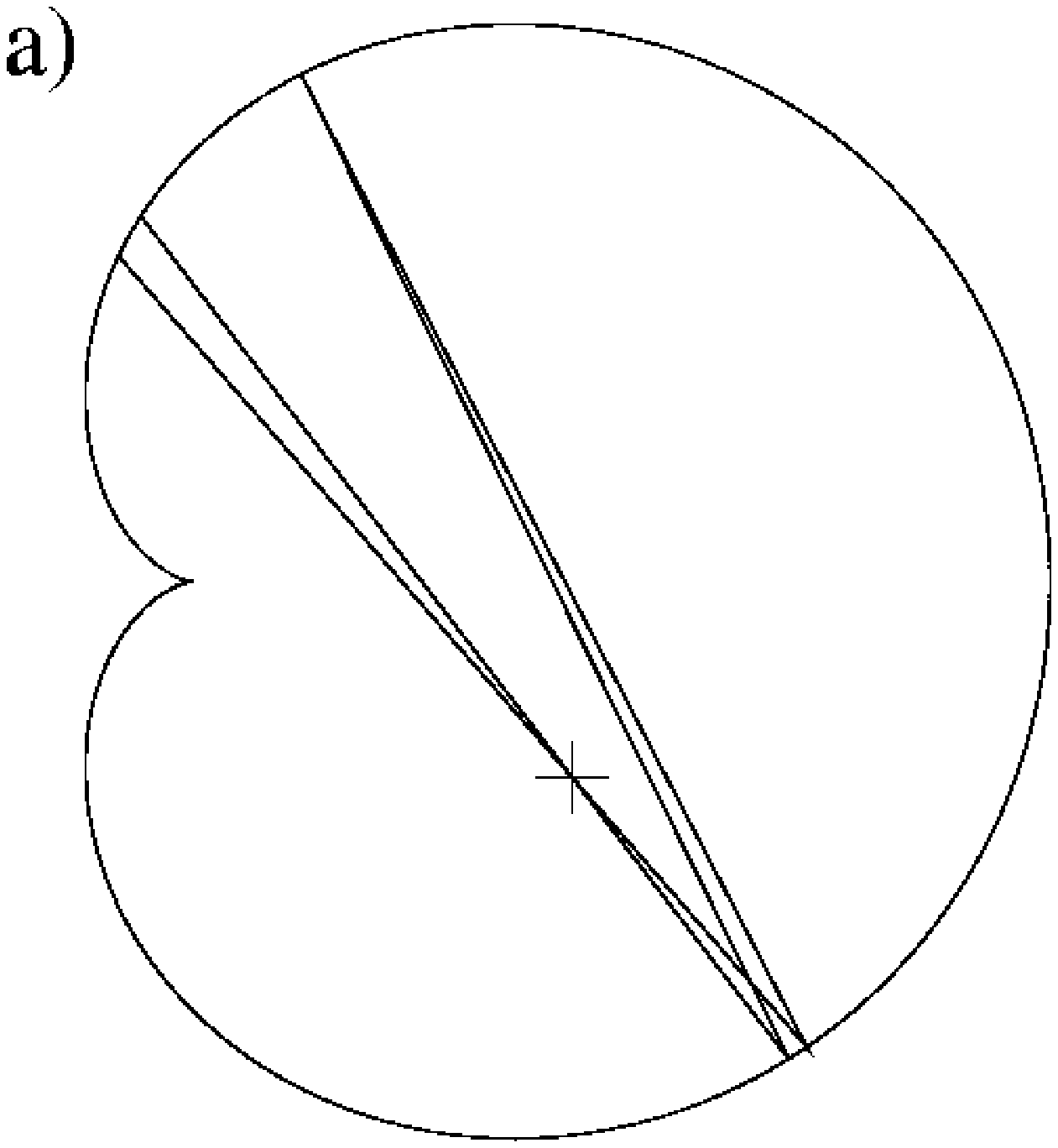}\hspace{1cm}
\includegraphics*[scale=0.27]{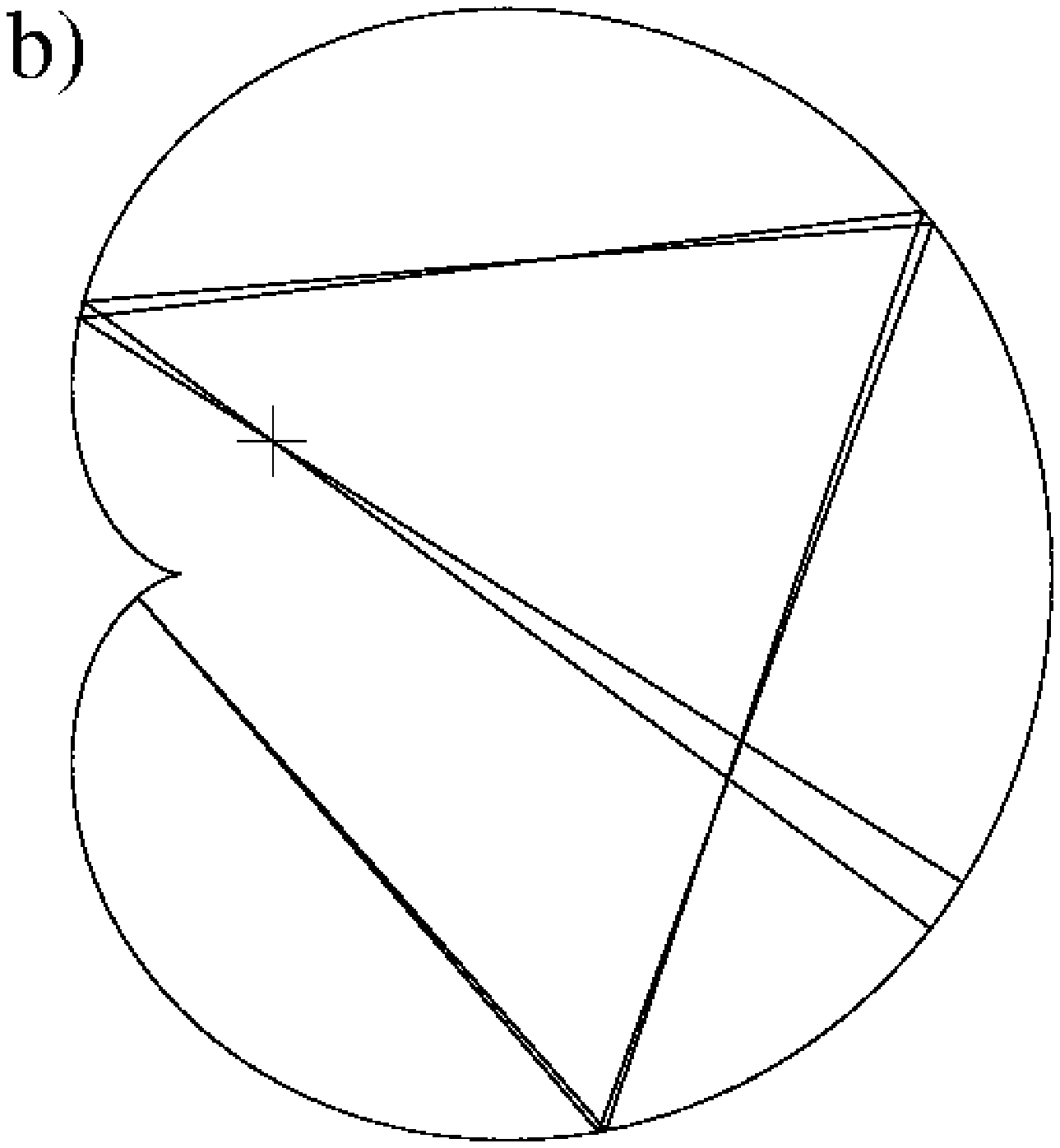}
\end{center}
\caption{Examples for crossings conjugate to a reflection in the cardioid billiard.}\label{fig:retracera}
\end{figure}

Examples for these loops are shown in Figure \ref{fig:retracera}. Here, a loop starts from the crossing, is reflected with an almost right angle and nearly retraces itself before crossing itself with a small angle. The locations of the crossing and the reflection are almost conjugate to each other, since the two traversals of the crossing limit a fan of trajectories with a small opening angle which gathers again at the reflection point.

These loops are so close to being time-reversal invariant that the ``partner'' formally determined by time reversal of one such loop coincides with the orbit itself. 
To show this, we use the notation introduced in Section 2 and let $L$ and $R$ denote the stability matrices of the left and right loop separated by the crossing (cp. Fig. \ref{fig:pair}).
For small crossing angles, the partner with time-reversed right loop has to fulfill the following linear system of equations (given in different form in \cite{siebrich}) for its traversals $\vec{x}_1'$, $\vec{x}_2'$ of a Poincar\'e section orthogonal to the first traversal of the crossing\footnote{We might as well consider the slightly more complicated Poincar\'e sections we had to choose in Section 2 for technical reasons.}
\begin{eqnarray}
\vec{x}_2'-\vec{x}_2 &=& L (\vec{x}_1'-\vec{x}_1)\nonumber\\
\vec{x}_1'-\T {\vec{x}_2} &=& R^\T (\vec{x}_2'-\T {\vec{x}_1}), \label{partnereq0}
\end{eqnarray}
where $R^\T=\T R^{-1} \T$ is the stability matrix of the time-reversed right loop. Note that in contrast to the reasoning in Section 2, here we do not require the orbit loops to be long. With $\delta\vec{x}_1=\vec{x}_1'-\vec{x}_1$, $\delta\vec{x}_2=\vec{x}_2'-\vec{x}_2$, $\delta \vec{x}= \T\vec{x}_2-\vec{x}_1=(0,|\vec{P}_1|\epsilon)$, (\ref{partnereq0}) simplifies to
\begin{eqnarray}
\delta\vec{x}_2&=&L\delta\vec{x}_1\nonumber\\
\delta\vec{x}_1-\delta\vec{x}&=&R^\T(\delta\vec{x}_2+\T\delta\vec{x}). \label{partnereq}
\end{eqnarray}
This system of equations in $\delta\vec{x}_1$ and $\delta\vec{x}_2$ is valid up to corrections quadratic in $\delta\vec{x}$, and obviously has exactly one solution. 
By going to higher orders, Spehner indeed rigorously showed that there is exactly one ``partner orbit'' \cite{spehner}. It is, among all periodic orbits, unambiguously singled out by fulfilling (\ref{partnereq}) in linear order. This ``partner orbit'' does however coincide with the initial orbit, if the right loop is smaller than the minimal loop time derived above. In this case, the separation between the two branches of the loop can be treated in a linear approximation. 
During the traversal time of the right loop $\vec{x}_2$ is carried into $\vec{x}_1$, and $\T\vec{x}_1$ is carried into $\T\vec{x}_2$. Since the separation between $\T\vec{x}_2$ and $\vec{x}_1$ can be treated in a linear approximation using the stability matrix of the right loop or its time-reversed, we obtain 
(up to quadratic order in $\delta\vec{x}$) 
\begin{eqnarray}
R^\T\T\delta\vec{x}=R^\T(\vec{x}_2-\T\vec{x}_1)=\vec{x}_1-\T\vec{x}_2=-\delta\vec{x}.
\end{eqnarray}
Comparing this to the second equation in (\ref{partnereq}), we see that (\ref{partnereq}) has the trivial solution $\delta\vec{x}_1=\delta\vec{x}_2=0$. Thus, the ``partner'' with time-reversed right loop coincides with the initial orbit. Conversely, if the left loop were shorter than the minimal loop time, that ``partner'' would coincide with the time reversal of the initial orbit.

In both cases, as pairs of identical or mutually time-reversed orbits are already included in the diagonal approximation, such crossings related to almost self-retracing loops give no off-diagonal contribution to the form factor. We conclude that in systems with conjugate points, the minimal loop time obtains a new meaning:
A partner orbit avoiding a given crossing exists only if both loops separated by that crossing exceed the minimal loop time. 
In the sequel, we will refer to crossings with an associated partner orbit as ``relevant'', and to the others as ``irrelevant'' crossings.
The crossing statistics derived above for systems without conjugate points immediately carries over to relevant crossings in systems with conjugate points. We refer to the Appendix for a discussion of the statistics of irrelevant crossings.  

For systems with symbolic dynamics, the irrelevance of almost self-retracing orbit loops can be shown even easier.
Each crossing divides the symbol sequence of the orbit in two parts $\sym{L}$ and $\sym{R}$ corresponding to the two loops\footnote{For simplicity, we assume here that each loop can unambiguously assigned a symbol sequence. Minor technical difficulties arise for symbols which effectively denote the crossing segments and not parts of a loop; they are dealt with in \cite{diplom}.}. The partner can be determined by reverting the symbol sequence of one loop in time.
An orbit with symbol sequence $\sym{LR}$ thus has a partner with symbol sequence $\sym{LR}^\T$, where the superscript ${}^\T$ denotes time reversal. The two are identical up to time reversal if one of the symbol sequences $\sym{L}$ or $\sym{R}$ is time-reversal invariant. This provides
a rigorous criterion to decide whether for a given crossing, there is an orbit avoiding the crossing or not.

\subsection{Numerical results}

We want to present numerical evidence that the distribution of relevant crossings in the cardioid indeed conforms to (\ref{np1}) and (\ref{pnp}). We looked for self-crossings in $8\times 10^6$ non-periodic orbits of length $L=250$ (which is long compared to the mean free path $1.851$ \cite{baeckerdullin} and the Lyapunov length $2.83$); for crossings with angles $<\frac{\pi}{10}$, even $4\times 10^8$ orbits were considered. In order to exclude irrelevant crossings from our statistics, we used the symbolic-dynamics criterion derived above.

Again, we have to deal with the effect of singularities on the crossing statistics. In the cardioid billiard, there are system-specific crossings related to the cusp, which will be discussed in more detail in the Appendix. Since for the corresponding orbits, even the applicability of the Gutzwiller trace formula is questionable, we find it necessary to also distinguish in our numerics between generic and cusp-related crossings. Like irrelevant crossings, the latter are excluded from our statistics using a symbolic-dynamics criterion derived in the Appendix.

\begin{figure}\begin{center}\includegraphics[scale=0.33]{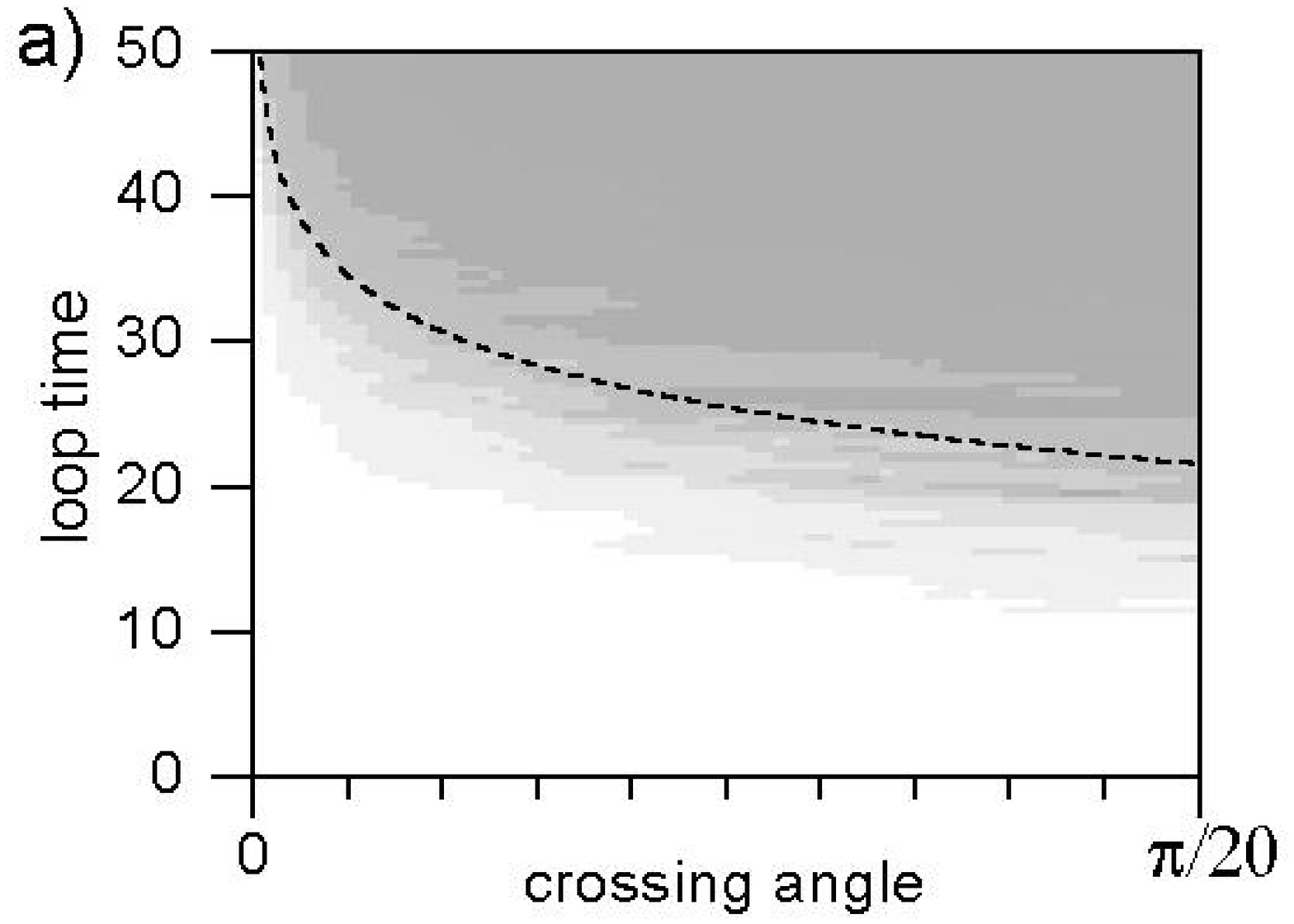}\hspace{1cm}\includegraphics[scale=0.33]{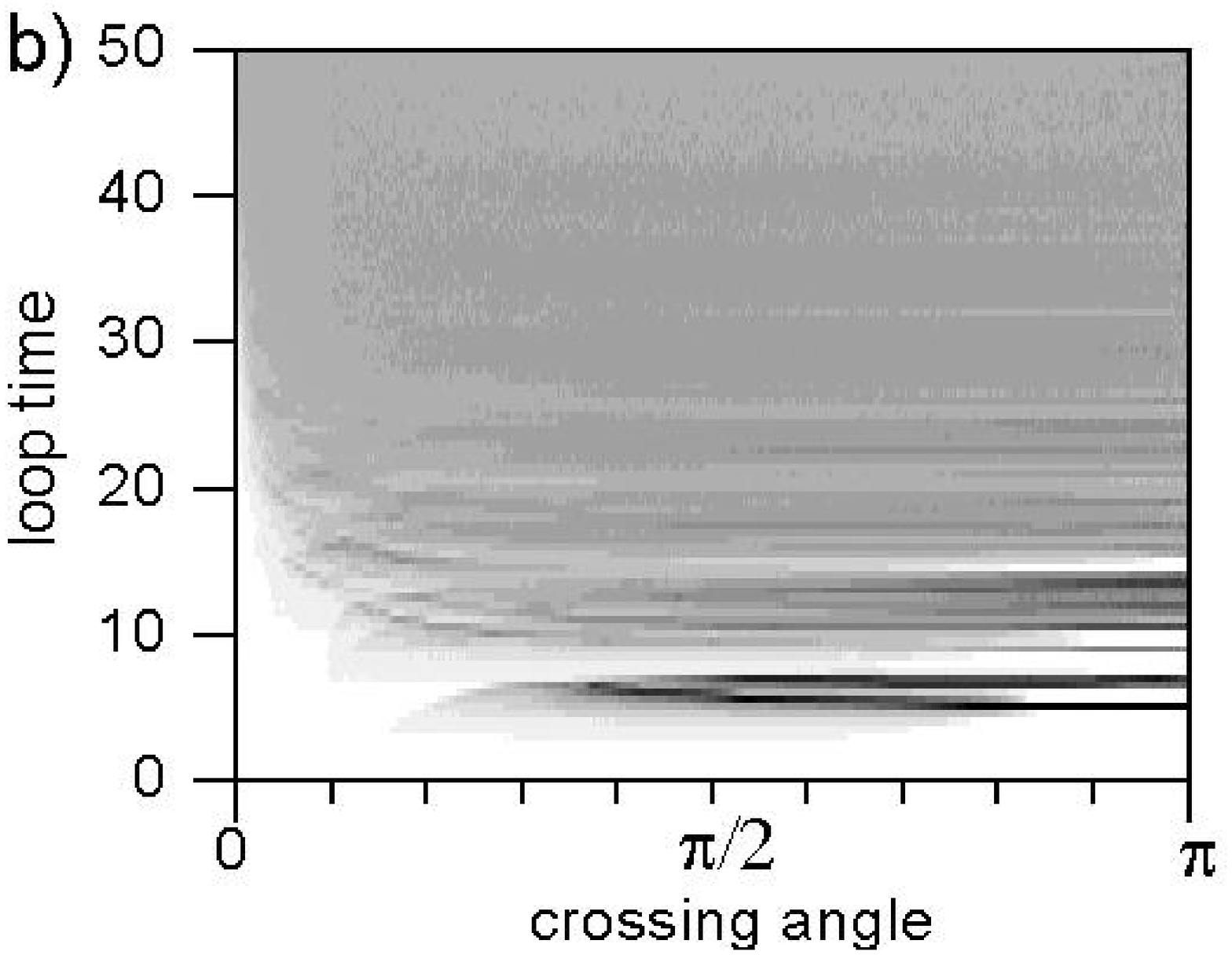}\hspace{1cm}\includegraphics[scale=0.165]{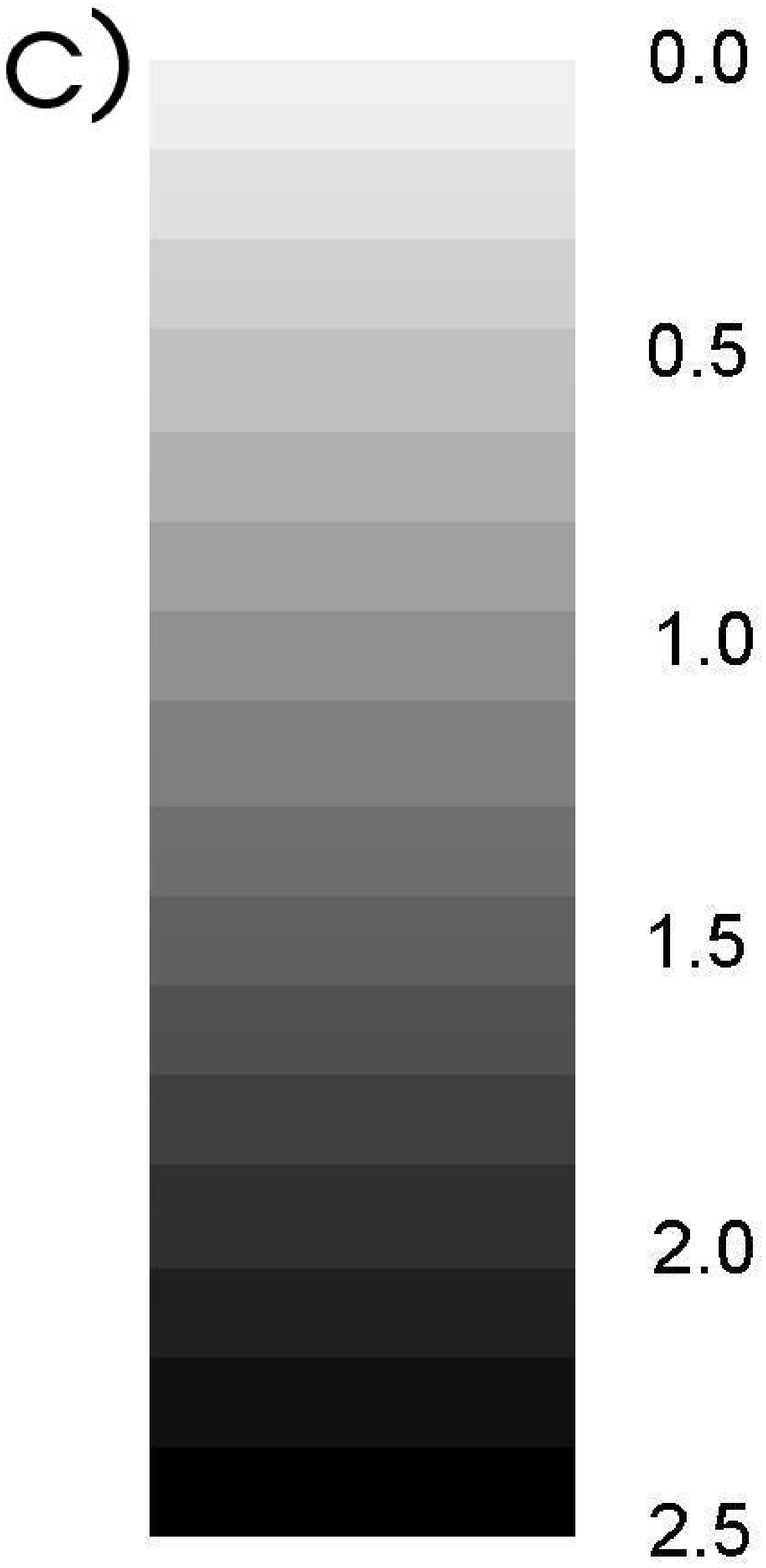}\end{center}\caption{Density plot of the combined density of loop times $t$ and crossing angles $\epsilon$ in the cardioid billiard: a) for $\epsilon<0.05\pi$, b) for all angles. Normalization as in Figure \ref{fig:diama}; the resulting scale is shown in c).
For small angles, we observe a threshold logarithmic in $\epsilon$, as indicated by a dashed line.}\label{fig:carda}\end{figure}

\begin{figure}
\begin{center}
\begin{tabular}{lp{0.5cm}l}
\includegraphics*[scale=0.19]{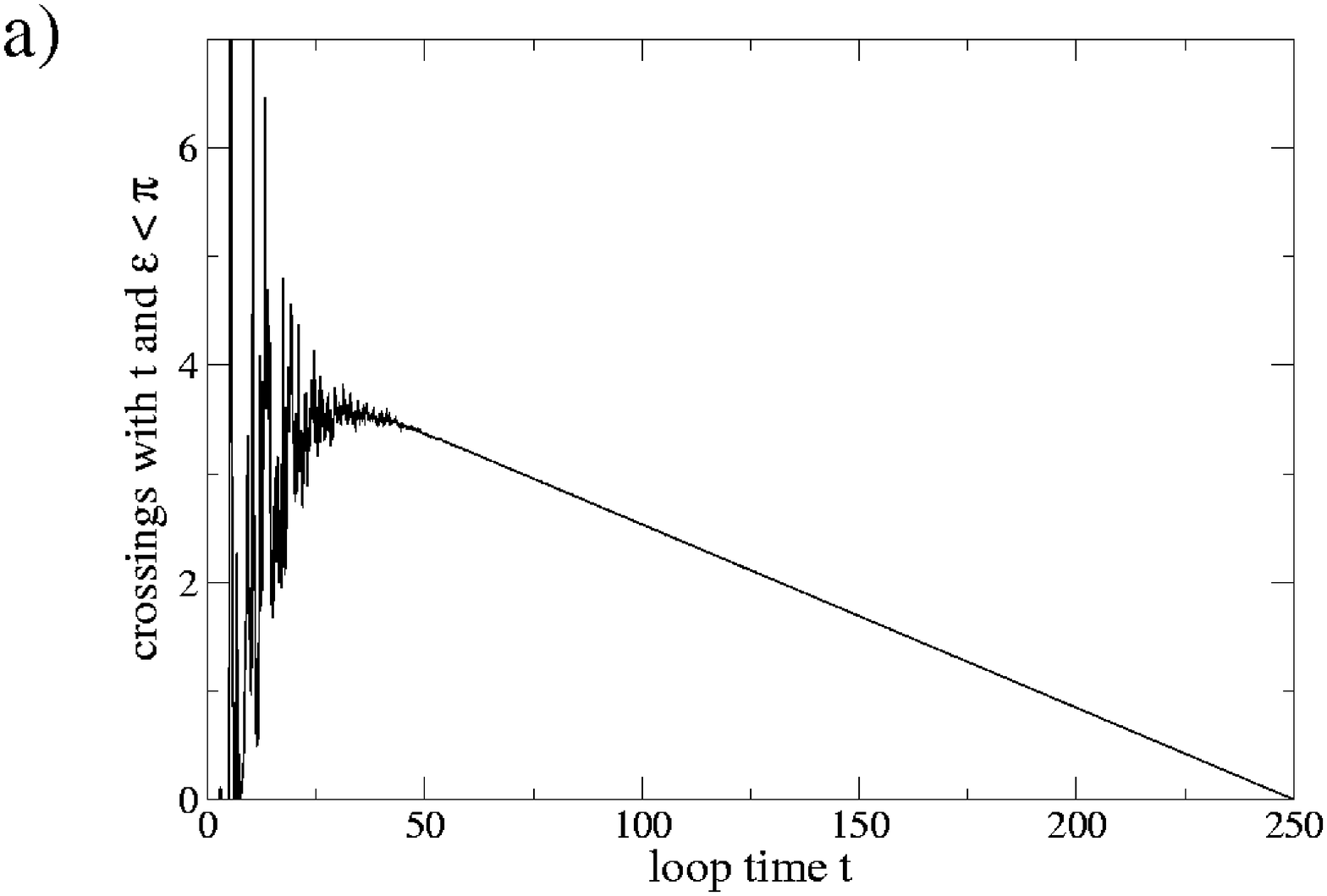}&&
\includegraphics*[scale=0.19]{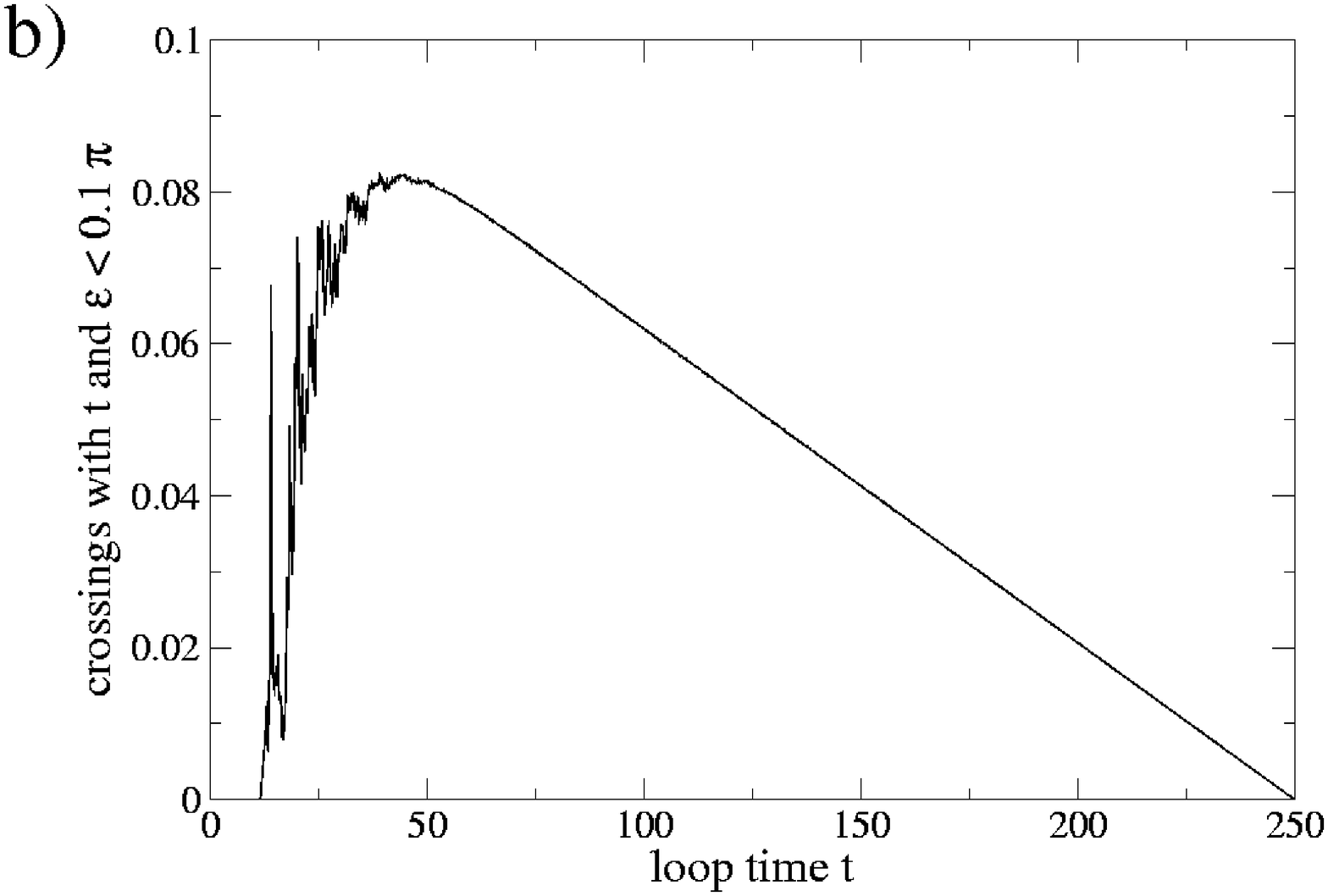}\\&&\\
\includegraphics*[scale=0.19]{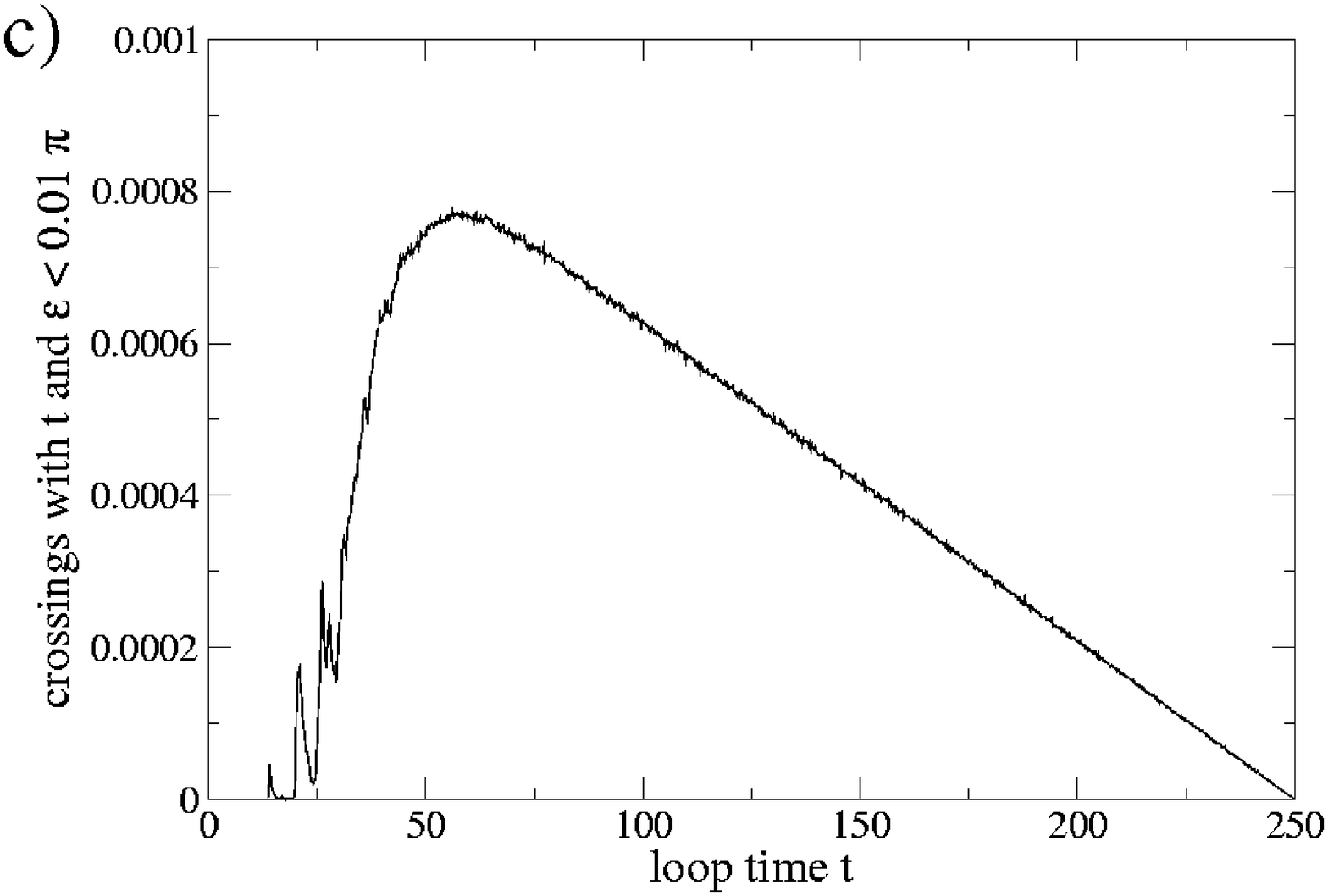}&&
\includegraphics*[scale=0.19]{c001.eps}
\end{tabular}
\end{center}
\caption{Statistics of loop times in non-periodic trajectories of traversal time $T=250$ in the cardioid billiard.
Depicted is the
density of crossings with loop time $t$ and crossing angle $<\epsilon_{\rm max}$ in a fraction of $\frac{1}{8}$ of the energy shell for a) $\epsilon_{\rm max}=\pi$, b) $\epsilon_{\rm max}=0.1\pi$, c) $\epsilon_{\rm max}=0.01\pi$, d) $\epsilon_{\rm max}=0.001\pi$.
Normalization as in Figure \ref{fig:l10}; the graphs are based on averages over $8\times 10^6$ orbits for large and $4\times 10^8$ orbits for small angles.
 As in Figure \ref{fig:l10}, due to the minimal loop time there is a gap for small $t$ which grows as $\epsilon_{\rm max}$ is reduced.}
\label{fig:c10}
\end{figure}

\begin{figure}
\begin{center}
\includegraphics*[scale=0.25]{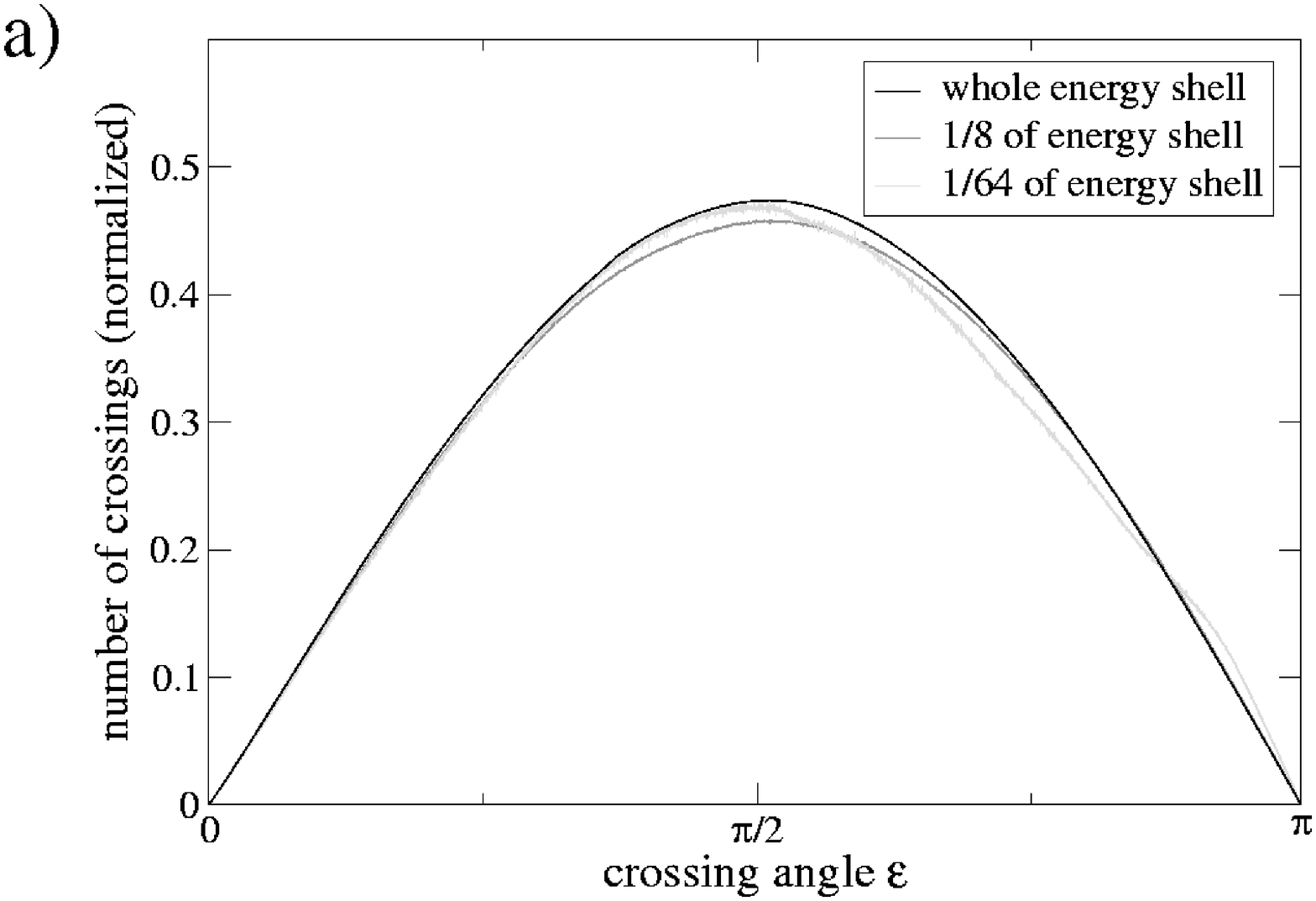}\hspace{1cm}
\includegraphics*[scale=0.25]{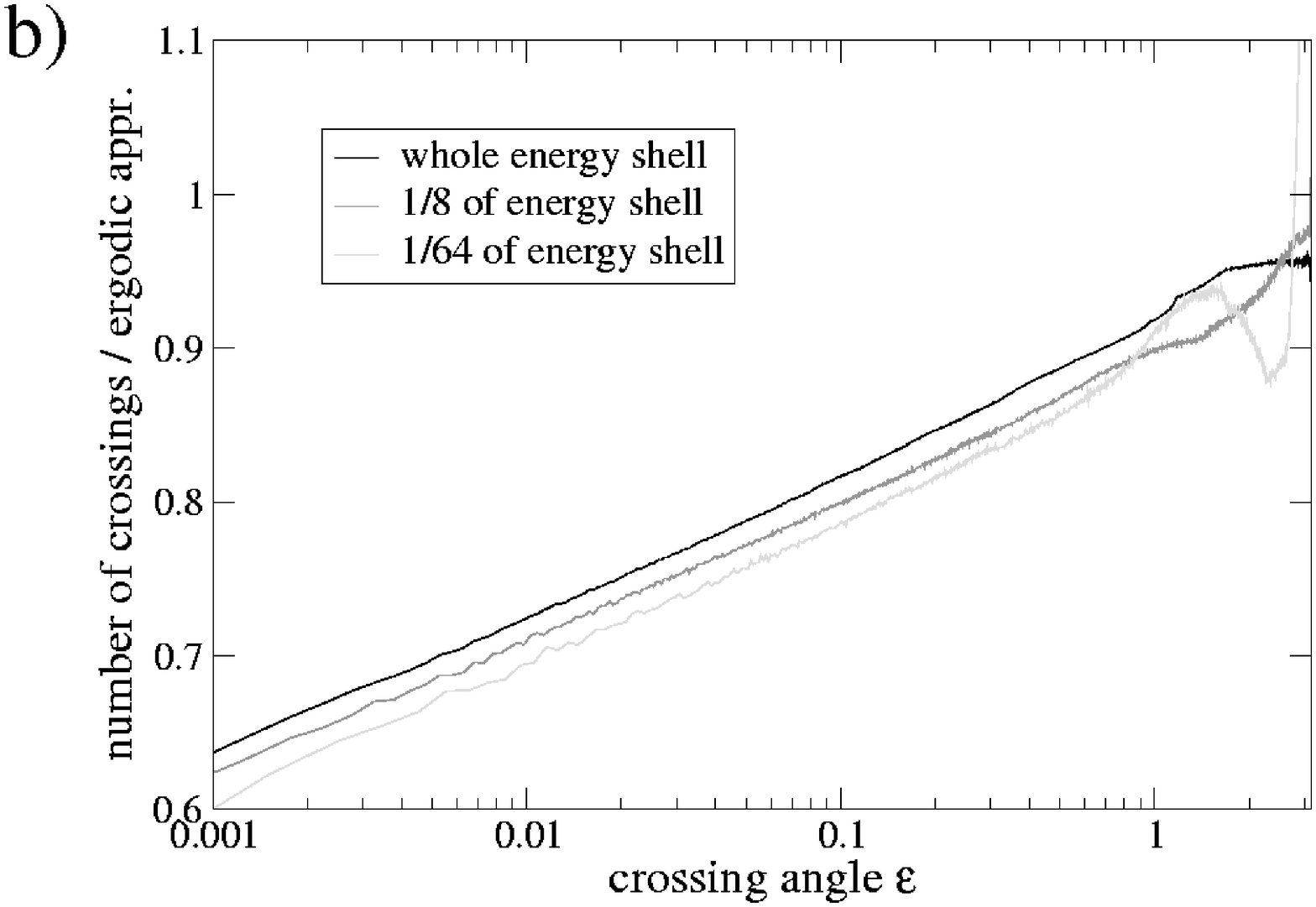}
\end{center}
\caption{a) Average of $P^{\rm np}(\vec{X},\epsilon|T)$ over the whole energy shell of the cardioid billiard and fractions of $\frac{1}{8}$ and $\frac{1}{64}$ thereof  (divided by $\frac{2\vec{P}^2T^2}{m|\Omega|^2}$ for normalization). b) The same quantities divided by $\frac{\sin\epsilon}{2}$ in a logarithmic plot. The logarithmic correction due to the minimal loop time becomes visible.}
\label{fig:cangle}
\end{figure}

Our numerical results for generic relevant crossings display a striking similarity to the case of the desymmetrized diamond billiard, thereby clearly supporting our point that the observed effects are universal.
Again, the combined density of crossing angles and loop times\footnote{In contrast to the following results, for this distribution it was still sufficient to include $3.5\times10^7$ trajectories.} reveals a minimal loop time depending logarithmitically on the angle, and some system-specific inhomogeneities close to that minimum (cp. Fig. \ref{fig:carda}).
Furthermore, the distribution of loop times corresponding to crossings in a given energy-shell bin and with a maximal angle $\epsilon_{\rm max}$ shows a gap for small angles (see Fig. \ref{fig:c10}). 
Finally, the density $P^{\rm np}(\vec{X},\epsilon|T)$ of generic relevant crossings shown in Figure \ref{fig:cangle} (averaged over the whole energy shell and over fractions of $\frac{1}{8}$ and $\frac{1}{64}$ thereof) agrees well with (\ref{pnp}). Based on the relative weight of the logarithm we obtain a fitting value for the Lyapunov exponent of $\lambda_{\rm fit}=0.352$ coinciding up to a minute numerical error with $\lambda=0.353$ \footnote{This value follows from the results of \cite{baeckerdullin} for the Kolmogorov-Sinai entropy of the billiard map $h_{\rm map}=\bar{l}\lambda=0.653$ and could be reproduced by averaging over the Lyapunov exponents of non-periodic orbits.}. 

\subsection{``Braids'' of crossings with common partner}

\begin{figure}
\begin{center}
\includegraphics*[scale=0.27]{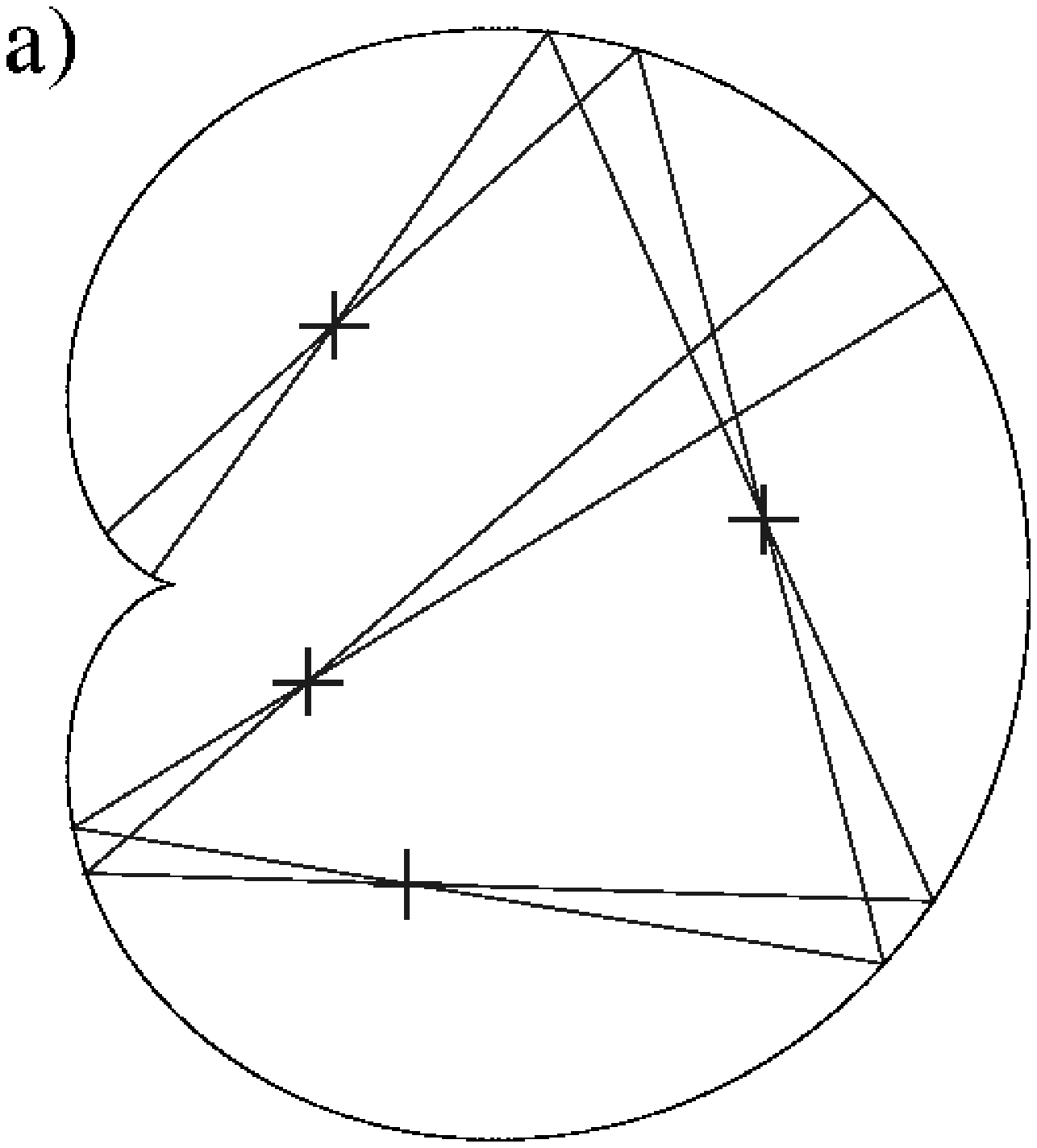}\hspace{1cm}
\includegraphics*[scale=0.27]{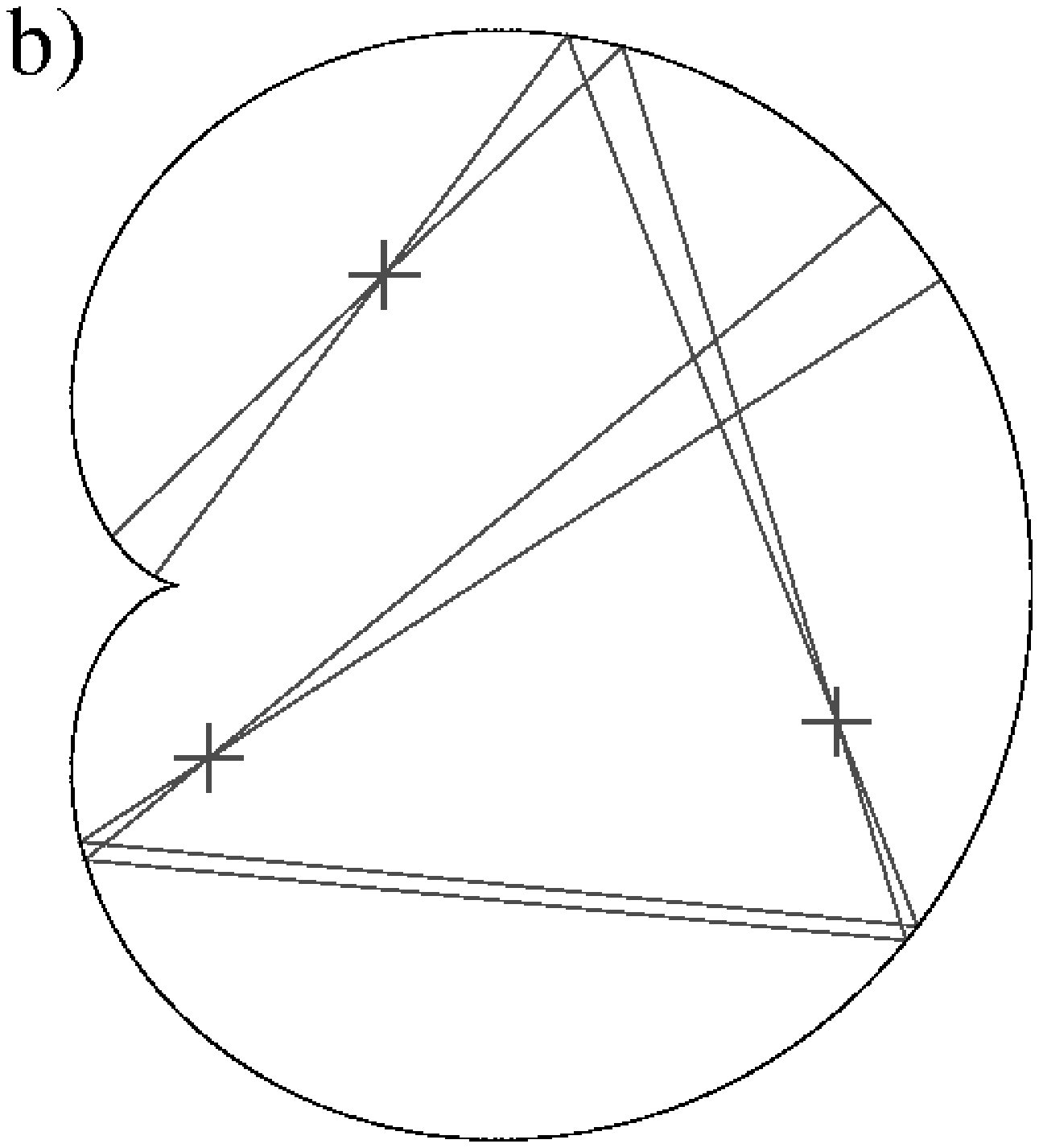}
\end{center}
\caption{a) Example for braid of crossings in mutually conjugate points in the cardioid billiard. We depict the central part of such a braid occurring in a periodic orbit. b) Crossings in the corresponding partner orbit. The number of crossings differs exactly by one.}\label{fig:family}
\end{figure}

In systems with conjugate points, two almost mutually time-reversed stretches of an orbit cross several times and thus have a whole family (``braid'') of small-angle crossings located close to mutually conjugate points. This is because two trajectories starting at the traversals of a given crossing will meet again in points conjugate to it, thereby forming new crossings. An example for such a braid of crossings in the cardioid is shown in Figure \ref{fig:family}a.

\begin{figure}\begin{center}\includegraphics*[scale=0.38]{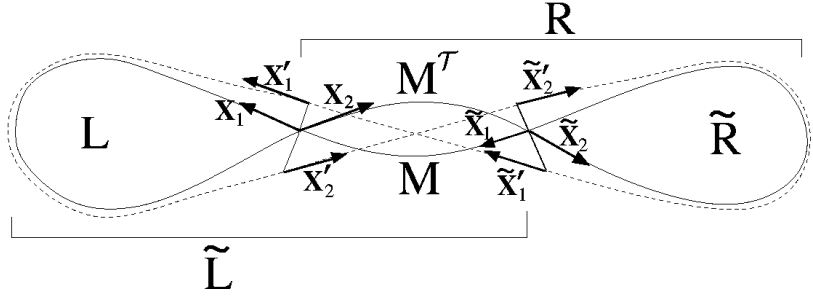}\end{center}\caption{Sketch of a Sieber-Richter pair with crossings in mutually conjugate points. The arrows denote phase-space points on the Poincar\'e sections defined in the text. $L$, $R$, $\tilde{L}$, $\tilde{R}$, and $M$ are stability matrices as described in the text.}\label{fig:familypair}\end{figure}

Interestingly, 
no matter which of these crossings one tries to avoid, one always obtains the same partner orbit.
To understand this, consider two crossings close to mutually conjugate points (cp. Fig. \ref{fig:familypair}). 
We introduce Poincar\'e sections orthogonal to one traversal $\vec{X}_1$ of the first and one traversal $\tilde{\vec{X}}_1$ of the second crossing. Furthermore, let $L$, $R$ ($\tilde{L}$, $\tilde{R}$) be the stability matrices of the loops separated by the first (second) crossing. If $M$ is the stability matrix describing the motion from $\tilde{\vec{X}}_1$ to $\vec{X}_1$, we have (in the notation introduced above)
\begin{eqnarray}
\tilde{L}&=&M^\T L M\nonumber\\
\tilde{R}&=&M^{-1}R(M^\T)^{-1}\nonumber\\
\delta\tilde{\vec{x}}&=&M^{-1}\delta\vec{x}.
\end{eqnarray}
The partner corresponding to the first crossing intersects the Poincar\'e section orthogonal to that crossing at $\vec{x}_i'=\vec{x}_i+\delta\vec{x}_i$, and a linear approximation shows that it intersects the Poincar\'e section orthogonal to the second crossing at $\tilde{\vec{x}}_i'=\tilde{\vec{x}}_i+\delta\tilde{\vec{x}}_i$ with
\begin{eqnarray}
\delta\tilde{\vec{x}}_1&=&M^{-1}\delta\vec{x}_1\nonumber\\
\delta\tilde{\vec{x}}_2&=&M^\T\delta\vec{x}_2.
\end{eqnarray}
It is easy to show that these $\delta\tilde{\vec{x}}_i$ also fulfill the system of equations determining the partner which corresponds to the second crossing ({\it i.e.} the ``tilded'' version of (\ref{partnereq}))\footnote{A similar result holds in a phase-space based treatment, as will be shown in \cite{spehner}.}. Consequently, for each braid of crossings close to mutually conjugate points and thus for any two approximately time-reversed stretches of an orbit there is just one partner orbit.

Again, we would like to point out how these findings translate to symbolic dynamics. The existence of two almost time-reversed stretches of an orbit is naturally reflected in its symbol sequence, because the stretches will have mutually time-reversed symbol sequences $\sym{Z}$ and $\sym{Z}^\T$. Thus, small-angle crossings appear in orbits with symbol sequences of the form
$\sym{lZrZ}^\T$ (where $\sym{l}$ and $\sym{r}$ are arbitrary symbol sequences) and occur between the stretches belonging to $\sym{Z}$ and $\sym{Z}^\T$ \cite{thepaper}. In case of conjugate points, there are several such crossings between these stretches. Any of them divides $\sym{Z}$ into two parts $\sym{Z_l}$ and $\sym{Z_r}$, and the whole symbol sequence into parts $\sym{L}=\sym{Z_l}^\T \sym{l Z_l}$ and $\sym{R}=\sym{Z_r r Z_r}^\T$ corresponding to the loops. The partner obtained by time reversal of the right loop has the symbol sequence
\begin{eqnarray}
\sym{L R}^\T=\sym{Z_l}^\T \sym{l} \sym{Z_l}(\sym{Z_r r Z_r}^\T)^\T=\sym{Z_l}^\T \sym{l Z_l Z_r r}^\T \sym{Z_r}^\T=\sym{lZr}^\T\sym{Z}^\T,
\end{eqnarray}
regardless of which crossing was chosen. 
Here, we used that time reversal of a symbol sequence inverts the ordering of its subsequences and reverts the subsequences in time, and that symbol sequences related by cyclic permutation are equivalent.
Thus, we see again that the orbit has only one partner for the whole braid.

This partner, too, contains a braid of crossings, as demonstrated for the example of the cardioid billiard in Figure \ref{fig:family}b. For both orbits within the pair, the crossing angles increase approximately exponentially towards the edges of the braids, where the two orbit stretches deviate more and more from being mutually time-reversed \cite{diplom}. The crossings of the partner orbit are slightly shifted compared to those in the original orbit (most strongly close to the center), and the numbers of crossings in both orbits differ by one, the orbit with larger action containing one crossing more.

\begin{figure}\begin{center}\includegraphics*[scale=0.4]{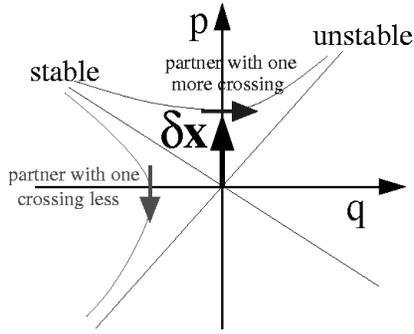}\end{center}\caption{Poincar\'e section orthogonal to one of two almost time-reversed orbit stretches with stable and unstable manifold and the vector $\delta\vec{x}$ pointing to the time reversal of the other stretch. Here, $\delta\vec{x}$ traverses the $p$-axis, {\it i.e.} the stretches cross in configuration space. For both Sieber-Richter partners, the asymptotic motion of $\delta\vec{x}$ with respect to the invariant manifolds is indicated by arrows.}\label{fig:poincare}\end{figure}

This observation, which will turn out to be crucial for the derivation of the form factor, will be proved in the sequel for general hyperbolic Hamiltonians of the form $H(\vec{Q},\vec{P})=\frac{\vec{P}^2}{2m}+V(\vec{Q})$. Note that it also trivially applies to systems without conjugate points, where the partner with larger action contains one crossing and the other none. Our proof relies on an argument of winding numbers. We follow one of the two almost time-reversed stretches of the orbit and study, in a Poincar\'e section orthogonal to the orbit, three quantities, the stable and unstable manifolds (which locally have the form of straight lines through the origin) and the (small!) phase-space vector $\delta\vec{x}$ pointing to the time reversal of the other orbit stretch (cp. Fig. \ref{fig:poincare}). As we move along the orbit, these lines and vectors rotate around the origin. Every time $\delta\vec{x}$ rotates through the $p$-axis, the orbit has a crossing. 
Note that for kinetic-plus-potential Hamiltonians the $p$-axis may only be traversed in clockwise direction, because because we have $\dot{q}=\frac{p}{m}>0$ in the upper and $<0$ in the lower half plane.

For the orbit with larger action, our formula for the action difference demands that whenever a crossing occurs we have $B_u(\vec{X})-B_s(\vec{X})>0$, {\it i.e.} the unstable manifold has a higher slope in the Poincar\'e section than the stable one. Thus for the partner with larger action, $\delta\vec{x}$ is located between the stable manifold (on the counter-clockwise side) and the unstable manifold (on the clockwise side), and for the partner with smaller action, the inverse is true. The motion of $\delta\vec{x}$ is given as a superposition of the rotation of invariant manifolds and a motion from the stable towards the unstable manifold (since the deviation between the orbit stretches first shrinks and then increases). We thus see that for the orbit with larger action, $\delta\vec{x}$ performs one more clockwise half-rotation around the orbit. Consequently, it crosses the $p$-axis once more and therefore the orbit with larger action contains one more crossing\footnote{This result holds even true in case of hard-wall reflections. Here, $\delta\vec{x}$ does not rotate continuously but jumps, and $q$ changes sign without traversal of the $p$-axis. However, as the two partners undergo the same number of hard-wall reflections, this does not affect the difference in numbers of crossings.}.

\section{Leading off-diagonal contribution to the form factor}
Once the crossing statistics and the action difference are known, we can now determine the contribution to the form factor arising from Sieber's and Richter's family of orbit pairs. 
It follows from (\ref{pairs}) that each pair of orbits of period $T=\tau T_H$ gives a contribution of $2 A_\gamma^2\cos\frac{\Delta S}{\hbar}$ to the form factor (where we neglect the difference in amplitude and period between the two partners).
However, we have to make sure that in spite of the one-to-one correspondence between crossings and orbit pairs being lost in general systems, each pair of orbits is counted only once.
The key to the solution is to formally assign to each crossing a contribution of $2 A_\gamma^2\cos\frac{\Delta S}{\hbar}\mbox{sign}(B_u(\vec{X})-B_s(\vec{X}))$. Since the partner with larger action and thus $B_u(\vec{X})-B_s(\vec{X})>0$ contains one more crossing that the partner with $B_u(\vec{X})-B_s(\vec{X})<0$, these formal contributions of all crossings in both partners add up to the correct value for the pair. We can thus sum over all Sieber-Richter pairs by summing over all orbits with period $T=\tau T_H$ and integrating over the locations $\vec{X}$ of the crossings on the energy shell $\Omega$ and over their angles $\epsilon$
\begin{eqnarray}
K_2(\tau)&=&\frac{2}{T_H}\left\langle\sum_\gamma A_\gamma^2\delta(T-T_\gamma)\right.
\int_\Omega d^3 X\mbox{sign}(B_u(\vec{X})-B_s(\vec{X}))
\left.\int_0^\pi d\epsilon\hspace{0.1cm}P(\vec{X},\epsilon|T)\cos\frac{\Delta S(\vec{X},\epsilon)}{\hbar}\right\rangle_{E,T}.
\end{eqnarray}
Note that each crossing is counted twice, since it is traversed at two mutually time-reversed phase-space points. As there are also two partner orbits with time-reversed left or right loop, no additional factor appears here.
The integral over the angle can be performed in a way similar to \cite{siebrich} using (\ref{actiondiff}), (\ref{minloop}) and (\ref{angle}) and yields in the semiclassical limit
$-\tau\frac{|B_u(\vec{X})-B_s(\vec{X})|}{2m\lambda|\Omega|}$.
This value only depends on the asymptotic action difference and angle distribution for $\epsilon\rightarrow 0$. Interestingly, the contribution of the ergodic approximation of the angle distribution vanishes, and the result is only due to its logarithmic correction originating from the minimal loop time. Applying the sum rule of Hannay and Ozorio de Almeida \cite{hoda} $\langle\sum_\gamma A_\gamma^2\delta(T-T_\gamma)\rangle_T=T$, we are led to
\begin{eqnarray}
K_2(\tau)
=-2\tau^2\frac{\langle B_u-B_s\rangle}{2m\lambda},\label{preff}
\end{eqnarray}
 where $\langle\ldots\rangle$ denotes an average over the energy shell.

Now, ergodic theory comes into play, relating the invariant manifolds to the Lyapunov exponent. 
The (positive) Lyapunov exponent of a hyperbolic two-freedom system can be expressed as the energy-shell average of the so-called local stretching rate
\cite{gaspard}\begin{eqnarray}
\chi(\vec{X})=\mbox{tr}\left[\frac{\partial^2 H}{\partial\vec{Q}\partial\vec{P}}+\frac{\partial^2 H}{\partial^2\vec{P}}C_u(\vec{X})\right].
\end{eqnarray}
Here, $C_u(\vec{X})$ is a matrix which relates the momentum and configuration-space components of
unstable deviations by $d\vec{P}=C_u(\vec{X})d\vec{Q}$. 
For systems with Hamiltonian $H(\vec{Q},\vec{P})=\frac{\vec{P}^2}{2m}+V(\vec{Q})$, this local stretching rate is proportional to $B_u(\vec{X})$. Evaluating the trace in coordinates orthogonal and parallel to the orbit, we see that $\chi(\vec{X})=\frac{1}{m}\hspace{0.05cm}\mbox{tr}\hspace{0.075cm}C_u(\vec{X})=\frac{1}{m}\hspace{0.025cm}B_u(\vec{X})$. Thus, we infer that $\langle B_u\rangle=m\lambda$.
Proofs for the special case of semi-dispersing billiards can also be found in \cite{bu}.
Due to $B_s(\T\vec{X})=-B_u(\vec{X})$, we also have $\langle B_s\rangle = -\langle B_u\rangle=-m\lambda$.
From this, we immediately obtain the universal leading off-diagonal contribution to the spectral form factor
\begin{eqnarray}
K_2(\tau)=-2\tau^2.
\end{eqnarray}

{\bf Acknowledgements:} I am grateful to Peter Braun, Fritz Haake, Stefan Heusler, and Dominique Spehner for close cooperation.
Furthermore, I want to thank Arnd B\"acker, Sven Gnutzmann, Gerhard Knieper, Klaus Richter, Holger Schanz, Martin Sieber, Lionel Sittler, Uzy Smilansky, Marko Turek, Carlos Viviescas, and Wen-ge Wang for enlightening discussions.

\section
{Appendix: More about crossings in the cardioid billiard}

\subsection*{Crossings in reflection-conjugate points}

It is instructive to see how the statistics of angles, locations and loop times of irrelevant crossings in ``reflection-conjugate'' points ({\it i.e.} points conjugate to the location of a self-retracing reflection) compares to the statistics of relevant crossings. We again consider the cardioid, where the distinction between these classes of crossings can be made with the help of symbolic dynamics.

\begin{figure}
\begin{center}
\begin{tabular}{lp{0.5cm}l}
\includegraphics*[scale=0.21]{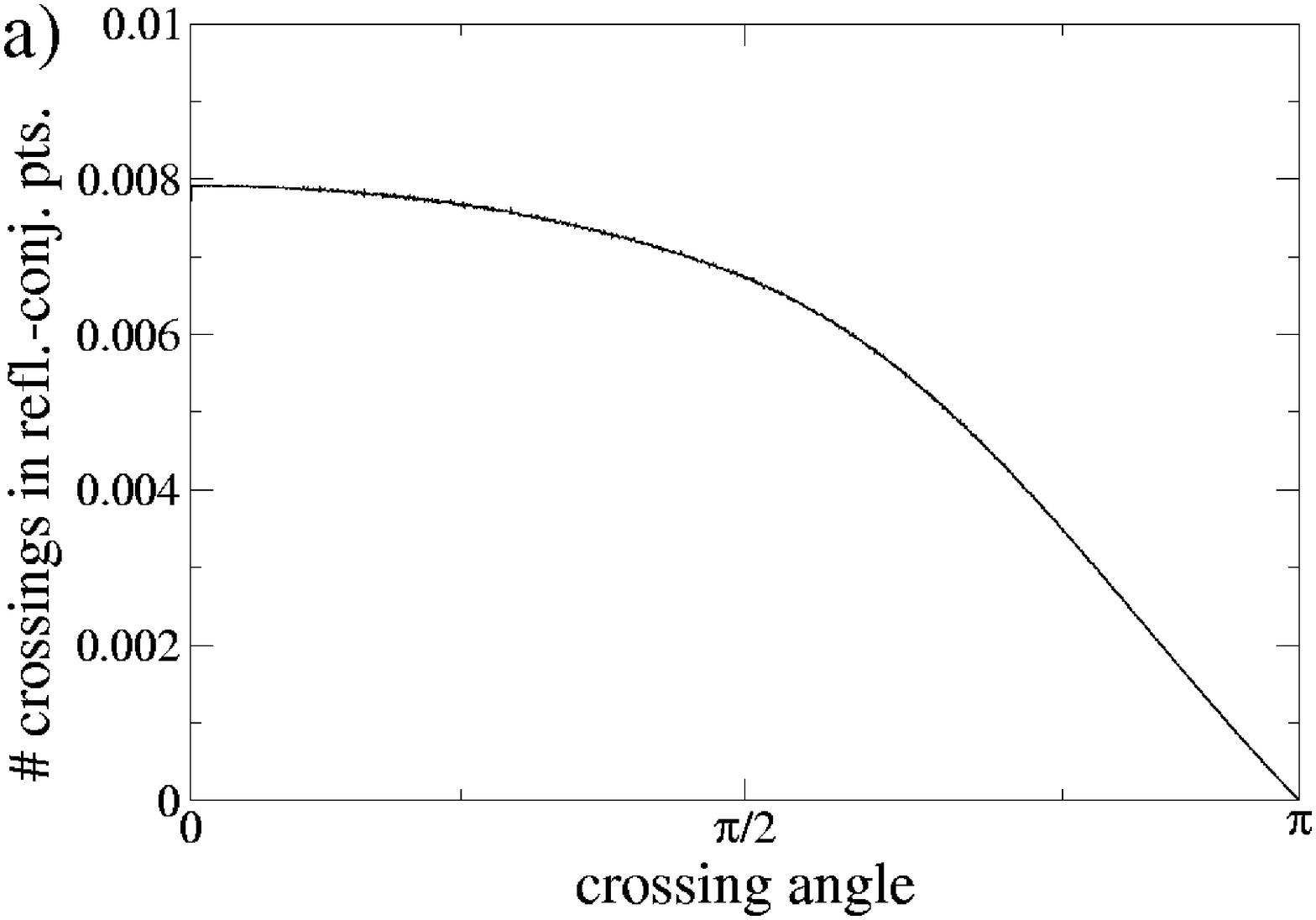}&&
\includegraphics*[scale=0.26]{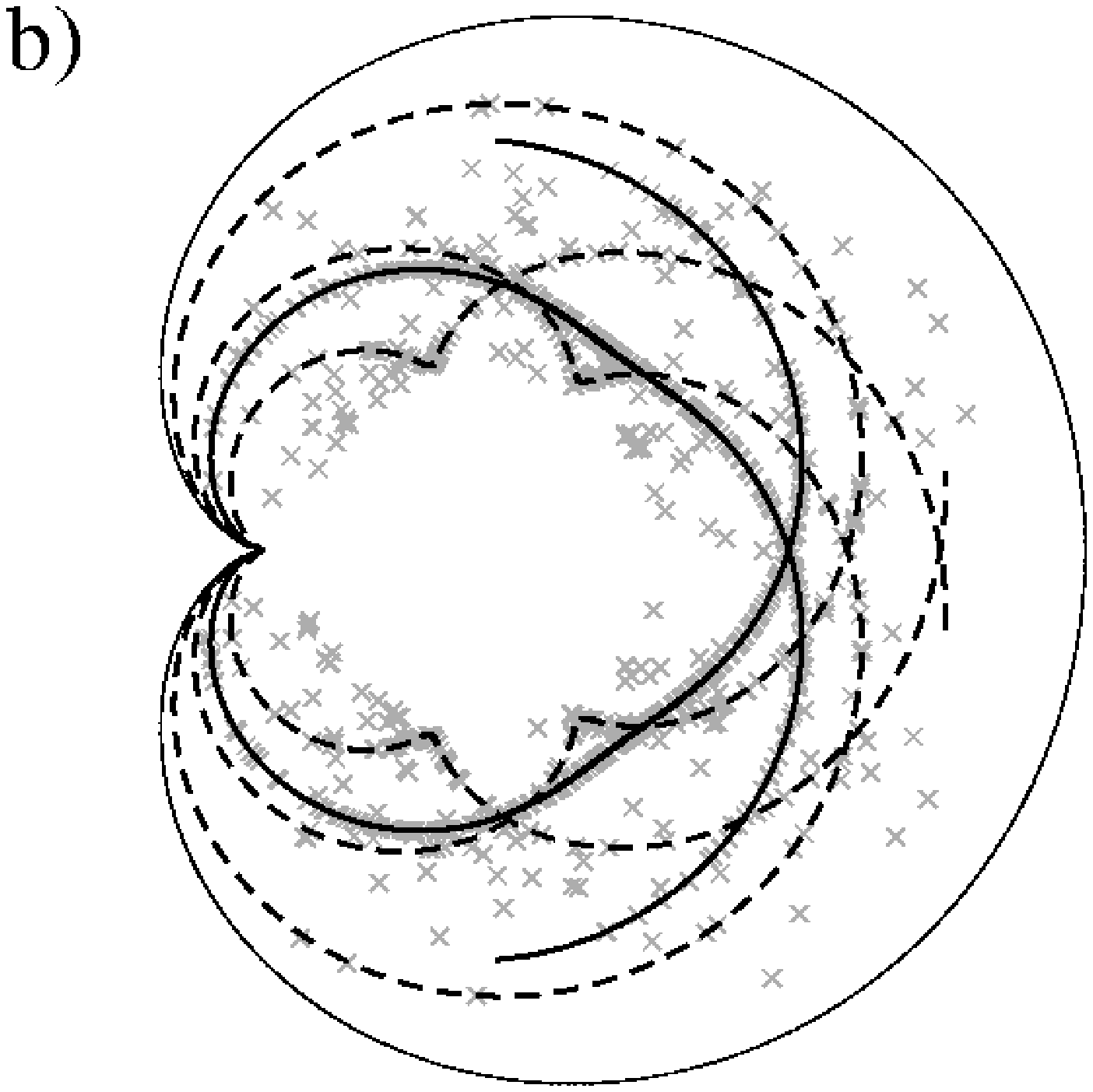}\\&&\\
\includegraphics*[scale=0.22]{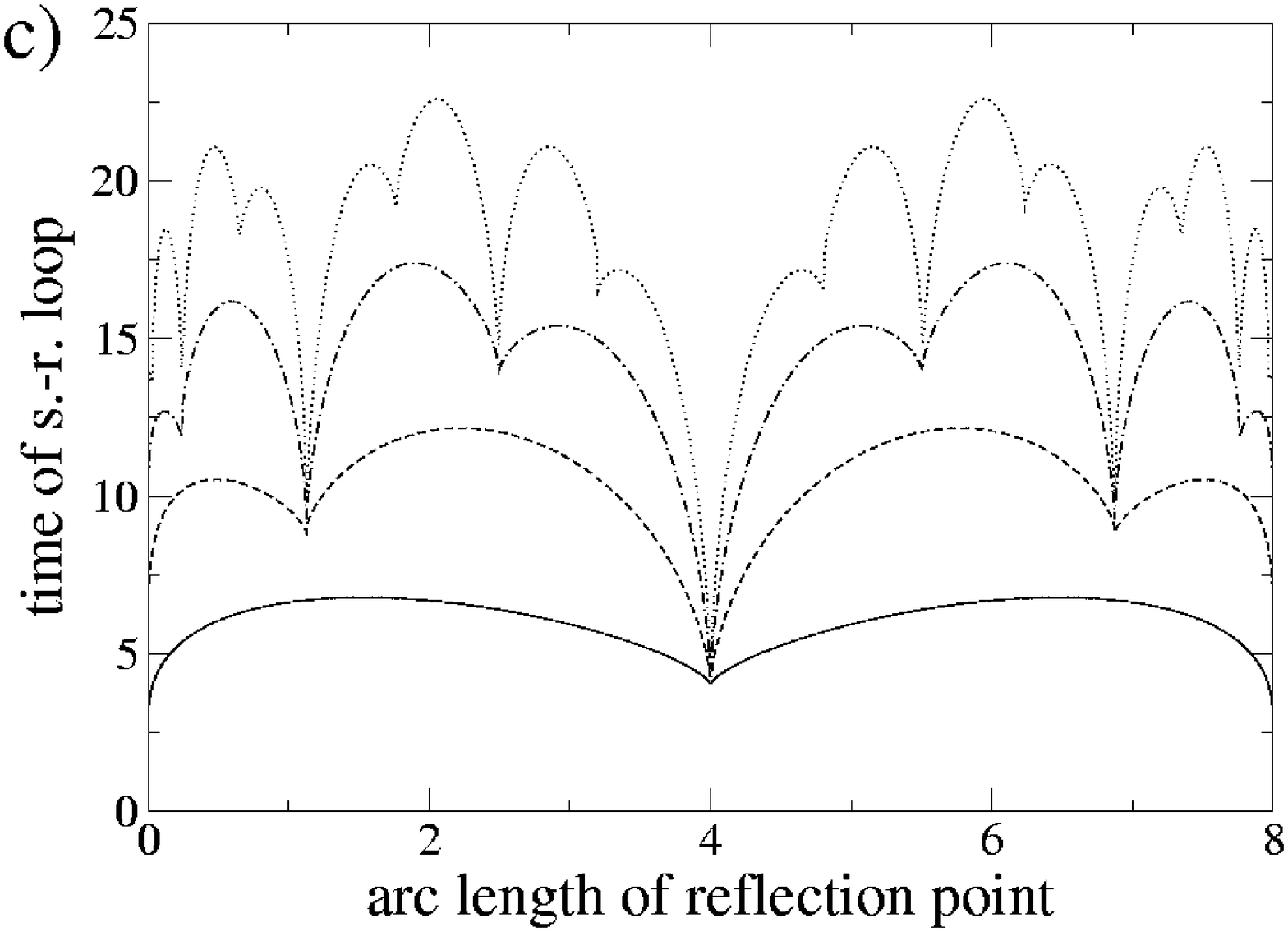}&&
\includegraphics*[scale=0.22]{retracer4.eps}
\end{tabular}
\end{center}
\caption{Statistics of irrelevant crossings with almost self-retracing loops in the cardioid billiard.
a) Density of angles $\epsilon$ corresponding to irrelevant crossings in non-periodic orbits of traversal time $T=250$ (divided by $\frac{2\vec{P}^2T^2}{m|\Omega|}$). 
b) Distribution in configuration space.
A random sample of irrelevant crossings found numerically is marked by crosses. They cluster close to the loci of reflection-conjugate points.
The locus of the first (second) conjugate point met after each self-retracing reflection is marked by a solid (dashed) curve.
c) Traversal times of the self-retracing loops, parametrized by the arc length of the reflection point on the boundary. Again, the different curves refer to the first, second, etc. conjugate point met after the reflection point.
d) Density of loop times for loops with crossing angle $<\frac{\pi}{10}$.}
\label{fig:retracer1}
\end{figure}

Numerically, we observe that the density of angles belonging to irrelevant crossings reaches a constant value for $\epsilon\rightarrow 0$ (see Fig. \ref{fig:retracer1}a), while for relevant crossings the corresponding density was shown to be proportional to $\sin\epsilon$ up to a logarithmic correction.
Thus, we see that in fact the vast majority of small-angle crossings are irrelevant.

In configuration space, irrelevant crossings are concentrated in the immediate vicinity of certain curves, the loci of reflection-conjugate points (see \ref{fig:retracer1}b). One can construct these curves by traveling along the boundary, starting trajectories in a direction perpendicular to the boundary, and determining points conjugate to these starting points using simple geometric optics. One such curve (sometimes divided into several parts) exists for the first, second, etc. conjugate points met after the starting point.

For each of these loci, the length (and thus time) of the trajectory from the boundary point to the crossings (corresponding to half of the length of the self-retracing loop) has several local maxima as shown in Figure \ref{fig:retracer1}c. A short calculation shows that these maxima are responsible for peaks in the loop time distribution which decay sharply towards larger and smoothly towards shorter time \cite{diplom} (see Fig. \ref{fig:retracer1}d).   

\subsection{Cusp-related crossings}

In this Appendix, we want to show that in the cardioid billiard, the minimal loop time is ignored by a system-specific class of crossings related to the singularity at the cusp. We will argue that for the evaluation of the form factor, these crossings have to be disregarded as well.

\begin{figure}\begin{center}\includegraphics*[scale=0.27]{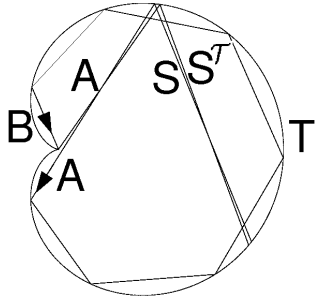}\end{center}\caption{An example for a cusp-related loop and its symbol sequence. The loop here has a symbol sequence of the form $\sym{SATBAS}^\T$ (with $\sym{S}=\sym{B}$ and $\sym{T}=\sym{B}^6$). 
Similar loops following the patterns
 $\sym{SBTABS}^\T$,
$\sym{SABTAS}^\T$, and $\sym{SBATBS}^\T$
are obtained by time reversal and reflection at the symmetry line.}\label{fig:plot}\end{figure}

An example for these crossings is shown in Figure (\ref{fig:plot}). For one of the two loops separated by the crossing, the two branches stay very near until being reflected on opposite sides of the cusp. Thus, the linear approximation for the separation between these branches breaks down before their separation reaches a typical phase-space scale. Then, in many cases, the loop closes by going around the boundary like in a whispering gallery. The length (and thus time) of the shortest such loops is approximately given by the circumference of the billiard.
In fact, they can be seen as deformations of finite orbits starting and ending at the cusp as introduced in \cite{bruuswhelan,baeckerdullin}.
In spite of their short time, the crossing angle can be arbitrarily small. So for these special loops, there is no minimal time depending logarithmitically on the angle.

However, they do give rise to pairs of orbits with similar action, as their symbol sequences are not time-reversal invariant.
Nevertheless, we have to disregard these crossings for a semiclassical evaluation of the form factor, as even the validity of the Gutzwiller formula for orbits coming so close to the cusp is questionable. 
In general, each time a periodic orbit comes closer to the boundary of a billiard than some length scale of the order of Planck length, so-called penumbra corrections to the trace formula play an important role, as shown in \cite{penumbra} for dispersing billiards.
The breakdown of the trace formula for families of orbits approaching the cusp of the cardioid 
is discussed in \cite{bruuswhelan,baeckerdiss}.

 Due to these problems, we have to distinguish in our numerics between generic and cusp-related crossings.
This distinction is done based on symbolic dynamics (compare Fig. \ref{fig:plot}). 
 The loop parts preceding the cusp are almost time-reversed with respect to each other and thus have symbol sequences $\sym{S}$ and $\sym{S}^\T$. In contrast, the part following the cusp is almost symmetric with respect to the symmetry line of the cardioid, {\it i.e.} its symbol sequence $\sym{T}$ has to be invariant under inverting the ordering of symbols \cite{baeckerdullin}. Usually, it undergoes a series of subsequent either clockwise or counter-clockwise reflections at the boundary, {\it i.e.} $\sym{T}$ just consists of  several identical symbols $\sym{A}$ or $\sym{B}$. In addition, for the symbols between $\sym{S}$, $\sym{T}$ and $\sym{S}^\T$ one has to take into account the effect of one branch being reflected close to the cusp and the other one narrowly avoiding it. These symbols can be read off from the example shown in Figure \ref{fig:plot}. Thus, we see that cusp-related crossings have symbol sequences of the form
$\sym{SATBAS}^\T$,
 $\sym{SBTABS}^\T$,
$\sym{SABTAS}^\T$, 
and $\sym{SBATBS}^\T$,
 where $\sym{S}$ is arbitrary and $\sym{T}$ invariant under inverting the ordering of symbols.

\end{document}